\documentclass[10pt,prd,twocolumn,
nofootinbib,
noshowkeys,
noshowpacs,
superscriptaddress,
floatfix
]{revtex4-2}
\usepackage{amsmath,amsfonts,amsthm,amssymb}
\usepackage[dvips]{graphics,graphicx}
\usepackage[usenames,dvipsnames]{color}
\definecolor{darkblue}{RGB}{0,0,196}
\definecolor{darkgreen}{RGB}{0,120,0}
\usepackage[colorlinks=true,linktocpage=true,linkcolor=darkblue,citecolor=red,urlcolor=darkblue]{hyperref}
\usepackage{cancel}
\usepackage{bbold}
\usepackage{multirow}
\usepackage{longtable}
\usepackage{color}
\usepackage[normalem]{ulem}
\usepackage{hyperref}
\usepackage{bigints}
\usepackage{xparse}
\usepackage{physics}
\usepackage{verbatim}
\usepackage{minibox}
\usepackage{comment}
\usepackage{appendix}
\usepackage{slashed}
\usepackage{marginnote}
\usepackage{graphicx}
\usepackage[nice]{nicefrac}
\usepackage{amsmath}
\usepackage{siunitx}
\usepackage{hepunits}
\usepackage{float}

\usepackage{stackengine,scalerel}
\newcommand\hstar[1]{\ThisStyle{\ensurestackMath{%
  \setbox0=\hbox{$\SavedStyle#1$}%
  \stackengine{0pt}{\copy0}{\kern.2\ht0\smash{\SavedStyle\star}}{O}{c}{F}{T}{S}}}}

\definecolor {darkgreen}{rgb}{0.2,0.7,0.2}

\begin{document}

\title{Effects of dark matter and magnetic field on neutron star properties in relativistic mean-field theory: A single-fluid approach}

\author{Neeshu Rani}
\email{neeshurani955@gmail.com}
\affiliation{Department of Physics, Birla Institute of Technology and Science Pilani,
Pilani Campus, Pilani, Rajasthan-333031, India}

\author{M. Mishra}
\email{madhukar@pilani.bits-pilani.ac.in}
\affiliation{Department of Physics, Birla Institute of Technology and Science Pilani,
Pilani Campus, Pilani, Rajasthan-333031, India}

\author{Deepak Kumar}
\email{deepakk@iiserbpr.ac.in}
\affiliation{Department of Physics, Indian Institute of Science Education and Research Berhampur, 760003,
India}

\author{Arpan Das}
\email{arpan.das@pilani.bits-pilani.ac.in}
\affiliation{Department of Physics, Birla Institute of Technology and Science Pilani,
Pilani Campus, Pilani, Rajasthan-333031, India}

\date{\today}

\begin{abstract}
Neutron stars, due to their extremely high matter density and strong magnetic field, provide the best environment for exploring new physics beyond the Standard Model of particle physics. In this work, we study the effect of pre-existing dark matter component and an internal magnetic field on the structural properties of neutron stars. We employed relativistic mean field theory based equations of state and used a single fluid approach for solving the Tolman-Oppenheimer-Volkoff (TOV) equation to compute properties like mass-radius, tidal deformability, compactness, and non-radial oscillation frequencies. We consider the following two scenarios for equation of state (EoS): (1) density-independent couplings along with non-linear interactions of mesons, and (2) density-dependent couplings, with only considering linear interactions for mesons. These mesons mediate the interactions between nucleonic constituents of a neutron star. In the dark matter sector we consider a massive fermionic dark matter which interacts with the nucleons through a Higgs portal interaction. We explore parameter regions for Fermi momentum of dark matter in the range $k_F = 0.01$ GeV - $0.06$ GeV, and two different values of the mass of fermionic dark matter, $M_\chi = 200$ GeV and $300$ GeV. We consider two values of the central magnetic field, $B_c = 7\times10^{17}$ Gauss, $9 \times 10^{17}$ Gauss, for a magnetized neutron star. Finally, we compare the theoretical predictions with the observed mass-radius and tidal deformability data of pulsars obtained from gravitational wave observations.
\end{abstract}

\maketitle

\section{Introduction}
Evidence for the existence of dark matter comes from the calculation of the mass of galaxy clusters by Zwicky and observation of the galaxy rotation curve by Vera Rubin ~\cite{deBlok:2001hbg,Bertone:2004pz,Bauer:2017qwy}. To date, the particle nature of dark matter remains unknown. Primordial Black holes (PBHs)~\cite{Kovetz:2017rvv,Loc:2024qbz,Dasgupta:2019cae,Carr:2016drx,Iguaz:2021irx,Xie:2024eug,
Jedamzik:2020omx} and Massive Astrophysical Compact Halo Objects (MACHOs) are the proposed heaviest astrophysical dark matter candidates. On the lower mass side, axions and/or Axion Like Particles (ALPs) are considered as dark matter candidates and commonly studied in neutron star environments~\cite{Lecce:2025dbz,Beznogov:2018fda,Fiorillo:2022piv,Dev:2023hax,Bhusal:2020bvx,
Sedrakian:2015krq,Zhang:2023vva, Kumar:2024abb}. Weakly interacting massive particles (WIMPs)~\cite{Kouvaris:2007ay,Quddus:2019ghy} are a group of well-motivated dark matter candidates that can arise in different extensions of the standard model that explain the relic abundance of dark matter reasonably well in the thermal freeze-out scenario. 

Several works have been carried out in the last few years assuming dark matter candidates either in the form of a pre-existing component in the core of neutron stars (NSs) or their formation in the core of NSs due to certain processes~\cite{Kumar:2026hoq}. For example, the cooling of a neutron star in particular, Cassiopea A neutron star, is explored in Ref.~\cite{Leinson:2021ety} employing emission of axions as dark matter particles. In Refs.~\cite{Quddus:2019ghy,Kumar:2026hoq}, the authors studied neutron star properties in the presence of WIMPs as constituents of the star by using mean field theory for the equation of state (EoS) and including the scalar (Higgs) portal and vector ($Z^\prime$) portal interactions between dark matter and normal matter. Because of no satisfactory results pertaining to the existence of WIMPs from either direct detection experiments or indirect probes, today's attention in searching for dark matter has been shifted towards the light dark matter below $1$ GeV. But the search for a heavy dark matter candidate is still ongoing. Compact objects, e.g., neutron stars, are so-called good natural astrophysical laboratories for searching for dark matter candidates~\cite{Rutherford:2022xeb,Sun:2023cqr, Pitz:2024xvh, Kumar:2024zzl, Das:2021yny,Liu:2024rix,Routaray:2022utr,Kumar:2025cro,Sen:2024yim}. Macroscopic properties of neutron stars, e.g., mass, radius, etc., are obtained by solving the Tolman-Oppenheimer-Volkoff equations. Dark matter can be present in the interior of neutron stars, e.g., due to accretion~\cite{DelPopolo:2020hel}. If dark matter is present in neutron stars, it can affect the properties of neutron stars~\cite{Kain:2021hpk,Kumar:2025ytm,Kumar:2024zzl,Kumar:2022amh,Acevedo:2024ttq,Bell:2023ysh,
Pal:2024afl,Shahrbaf:2025hsw,Cermeno:2017ejm,Kumar:2026hoq, Kumar:2024abb}.

Miao et al., in Ref.~\cite{Miao:2022rqj}, explored the parameter space for dark matter mass and self-interaction strength by studying the effect of dark matter on the pulse profile of stars using PSR J0030+0451 and PSR J0740+6620 X-ray observation data from NICER (Neutron star Interior Composition ExploreR). In Ref.~\cite{Lopes:2023uxi}, authors explored bosonic and fermionic dark matter in a strange quark star using the MIT bag model EoS. In Ref.~\cite{Thakur:2024ejl}, the authors computed the oscillation frequency for the non-radial mode of oscillations in a dark matter admixed neutron star (DMANS). In Ref.~\cite{Lee:2021yyn}, the authors studied the possibility for a $2.6 M_{\odot}$ object observed in the GW190814 event of being a dark matter admixed neutron star (DMANS) with bosonic dark matter. Various models have been employed for equations of state of normal and possible dark matter lying inside the core or crust. For example, in Ref.~\cite{Gusakov:2005it}, the authors studied cooling of a neutron star via neutrino emission through direct Urca process using the Akmal-Pandharipande-Ravenhal model, Ref.~\cite{Kumar:2025ytm} utilizes mean field EoS to study the same. Ref.~\cite{Abac:2021txj} considers three different EoSs for the crust part of neutron stars, named as Friedman–Pandharipande–Skyrme (FPS), Skyrme-Lyon (SLy), BSk19 EoS from the Brussels–Montreal Group, and the relativistic mean field approach has been used for the EoS of the core. The mean field framework includes the relativistic effects~\cite{Passarella:2025zqb} of highly dense matter present in neutron stars and can explain infinitely dense nuclear matter along with finite nuclei. Guha et al., in Ref.~\cite{Guha:2024pnn}, used models including density-dependent couplings for mesons in mean field EoS, along with constant coupling models exploring several EoS models. Furthermore, the strong internal and surface magnetic fields of neutron stars can affect their properties in nontrivial ways, which should be taken into account when exploring NS properties.

In the current work, we examine the properties of the neutron stars under the assumption that it is an admixture of baryonic matter and fermionic dark matter~\cite{Jyothilakshmi:2024xtl} in the presence of a non-vanishing internal magnetic field. The effect of the magnetic field on dark matter admixed neutron stars has been discussed in the literature~\cite{Parmar:2023zlg,Routaray:2024fcq}. However, in these studies, the modification of the TOV equation due to the presence of a magnetic field has not been considered. The magnetic field at the interior of a magnetized neutron star or magnetar can be as large as $10^{18}$ Gauss~\cite{Kaspi:2017fwg}. Such a large magnetic field in the interior can affect the energy of the charged
particles due to Landau quantization, and can modify the equation of state and the structural properties~\cite{Chatterjee:2018prm,Avancini:2017gck,Tolos:2016hhl,Franzon:2016iai,Franzon:2015sya}. Ideally, in the presence of a magnetic field, the neutron star structure can deviate from spherical symmetry and Tolman-Oppenheimer-Volkov (TOV) equations are no longer applicable for such an anisotropic  star~\cite{Chatterjee:2014qsa,Franzon:2015sya,Bowers:1974tgi,Bonazzola:1993zz}. Fundamentally, it is inconsistent to solve spherically symmetric equations for magnetized neutron star models, neglecting non-spherical deformation due to the presence of a magnetic field. But a relatively simpler approach has been developed in Ref.~\cite{Chatterjee:2018prm}, where one introduces a modified  TOV system by adding a magnetic field contribution to the energy density and a Lorentz factor term~\cite{Chatterjee:2018prm} in the hydrostatic equilibrium equation. This modified TOV equation can be used to study the macroscopic properties of neutron stars, qualitatively. In this study, we consider the approach discussed in Ref.~\cite{Chatterjee:2018prm} to examine the effect of the magnetic field on the properties of dark matter admixed neutron stars. We use here the modified version of publicly available code, TOVsolver~\cite{Tolman:1939jz, Oppenheimer:1939ne, Baym:1971pw, Hinderer:2009ca} for integrating TOV equations.

We compute here the mass-radius ($M-R$) sequences, tidal deformability ($\Lambda$), compactness ($C$) and non-radial oscillation frequency. We consider the interaction between dark matter and normal matter particles through the Higgs boson. In the relativistic mean field EoS, we include scalar ($\sigma$), vector ($\omega$) and isovector ($\rho$) mesons along with nucleons, leptons, dark matter and Higgs boson. The modified version of publicly available python code named CompactObject~\cite{Huang:2023grj, Huang:2024ewv, Huang:2024wig} for mean field EoSs is used by including the contribution arising from dark matter component. 

For dark matter, we assume uniform dark matter density throughout the neutron star with fermi momentum in the range $k_{f, DM} = 0.01-0.06$ GeV and two values for mass given as $M_{\chi} = 200$ GeV, $300$ GeV. We consider heavy dark matter because it has a higher capture rate in a neutron star as studied in Ref.~\cite{Bell:2020jou}. Hence, a large fraction of dark matter can accumulate inside stars and can significantly affect their properties. Ref.~\cite{Bell:2020jou} includes a detailed calculation for the capture rate, including relativistic effects for dark matter-nucleon scattering, by taking into account Pauli and multi-scattering effects. \\

The rest of the article is organized in the following manner. After this introduction, in Sec.~\ref{formalism} we discuss a single-fluid approach in the presence of dark matter component, magnetic field profile, and explain relativistic mean field theory in brief. Neutron star equation of state, mass-radius calculation, tidal deformability, compactness and non-radial oscillation mode frequencies have also been described in the same section. In Sec.~\ref{results_and_discussions}, we show the results and discussions. Finally, in Sec.~\ref{conclusion}, we conclude our main findings with a possible future outlook. 

\section{formalism}
\label{formalism}

In this section, we describe the single-fluid formalism for computing mass-radius ($M-R$) sequences for both magnetized and non-magnetized neutron stars consisting of normal nuclear matter plus dark matter (NSNDM). In our model of neutron stars including dark matter, we consider protons and neutrons as nucleonic constituents of the star represented by NM (nuclear matter). While the dark matter part is depicted by DM. Electrons and muons are also included as the leptonic sector of stars and are denoted by LS. 

\subsection{Single Fluid Formalism}

In a two-fluid picture, normal nuclear matter and dark matter components exist inside the neutron star as two non-interacting fluids, satisfying two separate TOV equations. In the two-fluid picture, NSNDM may have three possible configurations that have been explored in the literature, and are as follows~\cite{Miao:2022rqj}: 1) Neutron star with dark matter core, in which all the dark matter is present in the interior region of the star and can affect mass, pressure, energy density and temperature of the star. In this scenario, the total radius of the dark matter part is less than the total radius of the normal nuclear matter, which is the total radius of the star in this case, i.e., $R_{DM} < R_{NM} = R_{NS} $. 2) In the second case, a neutron star can have a dark matter sector in the form of a halo such that $R_{NM} < R_{DM} = R_{NS}$ and can alter the gravitational properties of stars like tidal deformability and gravitational lensing. 3) It can be a hybrid of dark matter and baryonic matter, i.e.,  when $R_{DM} = R_{NM} = R_{NS}$, may be called here as a hybrid star (HS). 

In our current investigation, we consider a single-fluid picture of a hybrid star (HS), where normal nuclear matter and dark matter can interact through a Higgs portal interaction.  Due to this self-interaction, we do not treat normal matter and dark matter as two distinct fluids in the TOV equation, but rather as a single fluid that is an admixture of the two. In order to calculate the macroscopic properties of a non-magnetized neutron star, we consider the spherically symmetric and static metric expressed as:
\begin{align}
ds^2 = -e^{\alpha(r)}dt^2 + e^{\beta(r)}dr^2 + r^2(d\theta^2 + \sin^2\theta \, d\phi^2),
\end{align}
with $g_{\mu\nu} = (-e^{\alpha(r)},e^{\beta(r)},r^2,r^2\sin^2\theta)$ being diagonal metric tensor in spherical polar coordinates $(t,r,\theta,\phi)$.
The solution of Einstein's equations in the interior and exterior regions of a star gives us the TOV equations and the Schwarzschild solutions, respectively. For the complete derivation of the TOV equation, the reader can refer to the paper~\cite{Das:2020ecp}. We consider $G=c=\hbar=1$ throughout the formulation, unless otherwise stated explicitly. The following are the TOV equations in a single fluid approach:
\begin{align}
& \frac{dP}{dr} = -\frac{m+4 \pi r^3P}{r^2(1-2m/r)}(\varepsilon+P) \nonumber\\
& \frac{dm}{dr} = 4 \pi r^2 \varepsilon 
\label{eq:2}
\end{align}   
where, $P(r) = P_{DM} + P_{NM} + P_{LS}$, $\varepsilon(r) = \varepsilon_{DM} + \varepsilon_{NM} + \varepsilon_{LS}$ are total pressure and total energy density of neutron star, respectively, with $P_{DM}$, $P_{NM}$, $P_{LS}$ ($\varepsilon_{DM}$, $\varepsilon_{NM}$, $\varepsilon_{LS}$) being the pressure (energy density) for dark matter, nucleonic matter and leptonic sector, respectively. Here $r$ is the radial distance from the center of the neutron star. These coupled differential equations are solved numerically by integrating them from the center of the star to the stellar surface, where the pressure vanishes, i.e., $P(r=R)=0$, with $R$ denoting the stellar radius. The boundary conditions imposed at the stellar center are $m(r=0)=0$ and $P(r=0)=P_c$, where $P_c$ is the prescribed central pressure. The total gravitational mass of the star is obtained from the mass function at the stellar surface, i.e., $M=m(R)$.

\subsection{Relativistic Mean Field Theory} 
\label{rmf_theory}

To solve the TOV equations, the equation of state (EoS), which relates the energy density and pressure, must be specified. EoS can also depend on baryon number density and temperature. In this calculation, we consider 
a zero-temperature EoS. We use the relativistic mean-field model to obtain the EoS, including two formalisms: (1) a non-linear model for meson self-interactions and density-independent couplings, and (2) a linear model for meson self-interactions but density-dependent couplings. In the density-independent frameworks, non-linear meson terms are included in the Lagrangian density of the nuclear matter in order to model the density dependence of the equation of state and symmetry energy~\cite{Boguta:1977xi,Mueller:1996pm,Steiner:2004fi,Todd-Rutel:2005yzo}. Similar effects can also be modeled without introducing non-linear mesonic terms, but incorporating density-dependent coupling coefficients~\cite{Typel:1999yq,Typel:2009sy,Lalazissis:2005de}. 

\subsection{Lagrangian density}
In our current work, we include $\sigma$, $\omega$ and $\rho$ mesons for writing the Lagrangian density of mean field EoS. Here, sigma ($\sigma$) and omega ($\omega$) fields act as mediators, respectively, for attractive and repulsive forces between nucleons. The rho ($\rho$) meson accounts for the isospin asymmetry of nucleons. Following are the Lagrangian density terms for nucleonic matter, meson fields and dark matter, respectively~\cite{Das:2020ecp,Das:2018frc}: 
\begin{align}
&  \mathcal{L}_{NM} = \bar\psi(i\gamma^\mu \partial_\mu - M_n - g_\sigma \sigma -g_\omega \gamma_\mu \omega^\mu \nonumber\\
&~~~~~~~~~~~~~~~~~~~~~~~~~~~~~~~~~~~~- g_\rho \gamma_\mu \vec{I}\cdot\vec{\rho}^{~\mu})\psi, \\
&    \mathcal{L}_{\sigma} = \frac{1}{2} \partial_\mu\sigma\partial^\mu\sigma - \frac{1}{2}m_\sigma^2\sigma^2 - \frac{1}{3}g_2\sigma^3 - \frac{1}{4}g_3\sigma^4, \\
&    \mathcal{L}_{\omega} =- \frac{1}{4}\Omega_{\mu\nu}\Omega^{\mu\nu} + \frac{1}{2}m_\omega^2\omega_\mu\omega^\mu, \\
&    \mathcal{L}_{\rho} =- \frac{1}{4}\vec{R}_{\mu\nu}\cdot \vec{R}^{\mu\nu} + \frac{1}{2} m_\rho^2\vec{\rho}_\mu\cdot\vec{\rho}^{~\mu},  \\
&    \mathcal{L}_{DM} = \bar\chi(i\gamma_\mu\partial^\mu - M_\chi)\chi.  \label{eq:lbm}
\end{align}
Here, $\chi$ is Dirac spinor for dark matter and $\bar{\chi} = \chi^\dagger\gamma^0$ is the adjoint spinor for dark matter. $g_\sigma$, $g_\omega$, $g_\rho$ represent couplings of nucleons with $\sigma$, $\omega$, $\rho$ fields, respectively, and $g_2, g_3$ are self-interaction associated couplings for sigma meson. $\psi$, $\bar{\psi}$ stand for nucleon doublet and its adjoint such that $\bar{\psi} = \psi^\dagger\gamma^0$ and: 
\begin{align}
    \psi = 
    \begin{pmatrix}
\psi_p \\
\psi_n
\end{pmatrix}, \quad
\psi^\dagger =
\begin{pmatrix}
\psi_p^* & \psi_n^*
\end{pmatrix}
\end{align}
where $\psi_p$ and $\psi_n$ represent the Dirac spinors for the proton and neutron, respectively. $\Omega_{\mu\nu} = \partial_\mu\omega_\nu - \partial_\nu\omega_\mu$ and $\vec{R}_{\mu\nu} = \partial_\mu\vec{\rho}_\nu - \partial_\nu\vec{\rho}_\mu$ are field strength tensors for $\omega$ and $\rho$ fields, respectively. $\vec I$ is the generator for the isospin of the proton and neutron. The interaction term between dark and nucleonic matter through the Higgs portal is given by:
\begin{align}
\mathcal{L}_H = \bar\chi yh\chi  + \frac{1}{2}\partial_\mu h \partial^\mu h - \frac{1}{2}M_h^2h^2 + \frac{fM_n}{v}\bar\psi h \psi,
\end{align}
where $h$ is the Higgs field, $y = 0.07$ , $M_h =125$ GeV  and $f = 0.35$ is Higgs-proton form factor. $v = 246$ GeV is vacuum expectation value for Higgs field~\cite{Cline:2013gha}. Higher order terms for Higgs fields are neglected, similar to what is employed in Ref.~\cite{Das:2018frc}, because only the linear order term is dominant in the mean field approach. Hence, the total Lagrangian density for the model considered here is given by: 
\begin{align}
    \mathcal{L}_T = \mathcal{L}_{NM} + \mathcal{L}_{DM} + \mathcal{L}_{\sigma} + \mathcal{L}_{\omega} + \mathcal{L}_{\rho} + \mathcal{L}_H.
    \label{densityinded_EoS}
\end{align}
In the mean field approximation, meson and Higgs fields are replaced by their expectation values, which are independent of space and time. Hence, the derivatives of these mean fields vanish. The mean field values of $\sigma$, $\omega$, $\rho$, $h$ are denoted as $\sigma_0$, $\omega_0$, $\rho_0$, $h_0$.

The only remaining unknown quantities that need to be fixed are the masses of nucleons and mesons, and the coupling strengths. In this work, we consider two frameworks for EoSs, density-independent \cite{Das:2020ecp} and density-dependent~\cite{Huang:2020cab,Malik:2022zol}. In the density-independent case, we consider the model described by the Lagrangian density given in Eq.~\eqref{densityinded_EoS}, with NL3 parameterization, where all the couplings and masses of meson fields and nucleons are given in Table (\ref{tb:1}).

We also consider density-dependent EoS, where the couplings of meson fields and nucleons depend on the baryon density ($\rho_B$). For this density-dependent EoS, we use a slightly different Lagrangian density than what is given in Eq. (\ref{densityinded_EoS}) for the density-independent scenario, with a positive sigma term as given below.~\cite{Huang:2020cab,Malik:2022zol,Char:2023fue}:
\begin{align}
   & \mathcal{L}_{NM} = \bar\psi(i\gamma_\mu \partial^\mu - M_n + g_\sigma \sigma -g_\omega \gamma_\mu \omega^\mu \nonumber\\
   & ~~~~~~~~~~~~~~~~~~~~~~~~~~~~~~- g_\rho \gamma_\mu \vec{I}\cdot\vec{\rho}^{~\mu})\psi.
   \label{densityded_EoS}
\end{align}
In a density-dependent coupling scenario, we consider $g_2 = g_3 = 0$, i.e., the non-linear self interactions for mesons vanish. We use two different parameterizations named DDMEX~\cite{Huang:2020cab} and DDBm~\cite{Malik:2022zol}. For the DDMEX model $g_{\sigma}$, $g_{\omega}$, and $g_{\rho}$ are defined as~\cite{Huang:2020cab},   
\begin{align}
& g_i(\rho) = g_ia_i\frac{1+b_i(x+d_i)^2}{1+c_i(x+d_i)^2} ; \quad i= \sigma,\omega , \nonumber\\
& g_{\rho}(\rho) = g_{\rho}e^{(-a_\rho(x-1))}. 
\label{eq:ddmex}
\end{align}
Parameters, $a_i, b_i, c_i, d_i$ for the density-dependent DDMEX model are given in Tables~(\ref{tb:2}). For the DDBm model $g_{\sigma}$, $g_{\omega}$, and $g_{\rho}$ are defined as~\cite{Malik:2022zol},   
\begin{align}
& g_i(\rho) = g_i~e^{(-(x^{a_i}-1))} ; \quad i= \sigma, \omega , \nonumber\\
& g_{\rho}(\rho) = g_{\rho}~e^{(-a_{\rho}(x-1))}. 
\label{eq:ddbm}
\end{align}
Here, $x=\rho_B/\rho_0$, $\rho_B$ is baryon density and $\rho_0$ is saturation density.  Parameters, $a_i$ for the DDBm model, are given in Tables~(\ref{tb:3}).

\begin{table*}[]
    \centering
    \resizebox{\textwidth}{!}{%
    \begin{tabular}{|c|c|c|c|c|c|c|c|c|}
        \hline
           $M_n$ (MeV) & $m_\sigma$ (MeV) & $m_\omega$ (MeV) & $m_\rho$ (MeV) & $g_\sigma$ & $g_\omega$ & $g_\rho$ & $g_2$ (fm$^{-1}$) & $g_3$\\
        \hline
        939& 508.194 & 782.501 & 763.000 & 10.217 & 12.868 & 4.474 & -10.431 & -28.885 \\
         \hline
          939&547.3327& 783 & 763 &  10.7067 & 13.3388 & 7.2380 & 0 & 0 \\
         \hline
         939& 550 & 783 & 763 & 9.180364 & 10.981329 & 7.652728 & 0 & 0 \\
         \hline
    \end{tabular}
    }
    \caption{Bare nucleon mass, meson masses, couplings in NL3 (first row)~\cite{Das:2020ecp}, DDMEX (second row)~\cite{Huang:2020cab}, DDBm (third row)~\cite{Malik:2022zol} parameterization, respectively.}
    \label{tb:1}
\end{table*} 

\begin{table*}[]
    \centering
    \begin{minipage}[t]{0.68\textwidth}
        \centering
        \resizebox{\textwidth}{!}{%
        \begin{tabular}{|c|c|c|c|c|c|c|c|c|}
        		\hline
            		$a_{\sigma}$  & $a_{\omega}$ &$a_{\rho}$ & $b_{\sigma}$  & $b_{\omega}$ & $c_{\sigma}$ &$c_{\omega}$   & $d_{\sigma}$ & $d_{\omega}$ \\
        		\hline
         	1.3970 & 1.3936&  0.6202 &1.3350 & 1.0191 & 2.0671 & 1.6060 & 0.4016 & 0.4556\\
         	\hline
    		\end{tabular}
        }
        \caption{Parameters for DDMEX parametrization~\cite{Huang:2020cab}.}
        \label{tb:2}
    \end{minipage}\hfill
    \begin{minipage}[t]{0.28\textwidth}
        \centering
        \resizebox{\textwidth}{!}{%
        \begin{tabular}{|c|c|c|c|}
        		\hline
         		$a_{\sigma}$  & $a_{\omega}$ &$a_{\rho}$  \\
        		\hline
         		0.086372 & 0.054065&  0.509147 \\
        		\hline
    		\end{tabular}
        }
        \caption{Parameters for DDBm parameterization~\cite{Malik:2022zol}.}
        \label{tb:3}
    \end{minipage}
\end{table*}

\subsection{Energy density and pressure}
\label{subsec:eos}
The non-magnetized equation of state (EoS), i.e., energy density ($\varepsilon$) and pressure ($P$), is obtained by calculating the energy-momentum tensor ($T^{\mu\nu}$) for the Lagrangian density given in Eqs.~\eqref{densityinded_EoS} and \eqref{densityded_EoS}~\cite{Das:2018frc}. The energy density and isotropic pressure are given as:  
\begin{align}
    \varepsilon = T^{00},~~~P= \frac{1}{3}\sum_{i=x,y,z}T^{ii}.
\end{align}
Here,
\begin{align}
&    \varepsilon = \sum_{i= n,p, \chi, e,\mu}\bigg[\frac{1}{8\pi^2} \bigg(k_{f,i} E_{f,i}^3 + k_{f,i}^3 E_{f,i} \nonumber\\
& ~~~~~~~~~~~~~~~~~~~~~~~ - \log\bigg(\frac{k_{f,i}+E_{f,i}}{M_{eff,i}}\bigg) M_{eff,i}^4\bigg)\bigg]\nonumber \\
& ~~~~~~   + \frac{1}{2}m_{\sigma}^2\sigma_0^2+ \frac{1}{3}g_2\sigma_0^3+ \frac{1}{4}g_3\sigma_0^4
    + \frac{1}{2}m_{\omega}^2\omega_0^2\nonumber\\
    & ~~~~~~~~~~~~~~~~~~~~~~~~~~~~~~ + \frac{1}{2}m_{\rho}^2 \rho_{0}^2 + \frac{1}{2} M_h^2 h_0^2,
\end{align}
and,
\begin{align}
  &  P = \sum_{i= n,p, \chi, e,\mu}\bigg[\frac{1}{12\pi^2}~ (1.5 M_{eff,i}^4\tanh^{-1}\bigg(k_{f,i}/E_{f,i}\bigg) \nonumber\\
  & ~~~~~~~~~~~~~~~ - 1.5~k_{f,i}M_{eff,i}^2 E_{f,i} + k_{f,i}^3 E_{f,i}\bigg)\bigg] \nonumber\\
  & ~~~~~~~ - \frac{1}{2} m_{\sigma}^2 \sigma_0^2  - \frac{1}{3} g_2 \sigma_0^3 - \frac{1}{4} g_3 \sigma_0^4
    + \frac{1}{2} m_{\omega}^2 \omega_0^2 \nonumber\\
  & ~~~~~~~~  + \frac{1}{2} m_{\rho}^2 \rho_{0}^2 - \frac{1}{2} M_h^2 h_0^2.
\end{align}
Here, the sum includes contributions for proton ($p$), neutron $(n)$, dark matter ($\chi$), electron ($e$) and muon ($\mu$). $k_{f, i}, E_{f, i}$ and $M_{eff, i}$ are Fermi momentum, Fermi energy and effective mass for i-th constituent, $i \in p, n, \chi, e, \mu$, respectively as given in Ref.~\cite{Das:2018frc}. Also, the effective masses of electrons and muons are the same as their bare masses (medium-independent mass) because medium dynamics only affect the masses of nucleons and the dark matter particle. These equations have been solved by using the beta equilibrium condition (for beta decay of nucleons in terms of chemical potentials of nucleons and leptons) and the charge neutrality condition expressed below:
\begin{align}
    \mu_n = \mu_p + \mu_e; \qquad
    \mu_e = \mu_\mu; \qquad
    \rho_p = \rho_e + \rho_\mu,
\end{align}
where $\mu_i$ and $\rho_i$ are chemical potentials and number densities, respectively, for the i-th species. For density-dependent EoS, the chemical potential and Fermi energy relation include a few extra terms in comparison to the density-independent case. Extra terms arise due to the fact that coupling constants, $g_{\rho}$, $g_{\sigma}$, $g_{\omega}$, are density dependent. Therefore, $\mu_i = \frac{\partial \varepsilon}{\partial\rho_i}$ contains the derivative of couplings with respect to baryon density~\cite{Huang:2020cab, Malik:2022zol}.

\subsection{Modified TOV equation for radial Magnetic field profile}

Magnetic field introduces anisotropy, and applying the TOV equation to get the structure of such a star would be inconsistent. Therefore, one should implement the hydrostatic equations for an anisotropic system, which is rather involved.  On the other hand, to qualitatively describe the effects of the magnetic field on neutron star properties, a modified TOV approach has been developed in Ref.~\cite{Chatterjee:2018prm}, 
\begin{align}
& \frac{dP}{dr} 
= -c^2\left(\varepsilon + \frac{B^{2}(r)}{2\mu_{0}c^2} + \frac{P}{c^{2}}\right)
\left(\frac{d\Phi}{dr} - L(r)\right),\label{dpdrmagnetic}\\
& \frac{d\Phi}{dr} 
= \frac{G\left(m + 4\pi\frac{P}{c^{2}}r^3\right)}{r(rc^2 - 2Gm)}, \\
& \frac{dm}{dr} 
= 4\pi r^{2}\left(\varepsilon + \frac{B^{2}(r)}{2\mu_{0}c^2}\right),\label{dmdrmagnetic}\\
& \frac{L(r)}{10^{-41}} = - B_c^2 \left[3.8\,x - 8.1\,x^3 + 1.6\,x^5 + 2.3\,x^7 \right], \\
& B(r) = B_c \left[ 1 - 1.6\,x^2 - x^4 + 4.2\,x^6 - 2.4\,x^8 \right].
\end{align}
Here, $\varepsilon$ is the density of the radial mass and $x=r/\bar{r}$, $\bar{r}$ is the mean radius arising from deformation produced by the strong internal magnetic field~\cite{Chatterjee:2018prm}. In the current work, we consider $\bar{r}=14$ km~\cite{Yadav:2022yqa} (slightly larger than the actual radius of the NSs). A radial distance dependent magnetic field profile is used, whose effect is included through the magnetic energy density term and a Lorentz factor $L(r)$ in $cm/sec^2$. 
In Eqs.~\eqref{dpdrmagnetic}-\eqref{dmdrmagnetic}, $G$, $c$, and $\mu_0$ appear explicitly. However, in our calculation, we consider $G=c=\hbar=1$, and convert the energy density of the magnetic field from Gauss$^2$ into GeV$^4$ units. To obtain the energy density in Gauss$^2$ units in Eqs.~\eqref{dpdrmagnetic}, and \eqref{dmdrmagnetic}, we use a Gaussian system of units, i,e., $\mu_0=4\pi$ and replace $B^2/2\mu_0$ with $B^2/(8\pi)$~\cite{Strickland:2012vu}. The magnetic field profile is parameterized using the parameter, $\bar{r}$ i.e., mean radius of the star, and $r$ represents the internal radial distance from the center used in a spherically symmetric metric as mentioned in~\cite{Chatterjee:2018prm}. In the limit $B_c\rightarrow 0$, Eqs.~\eqref{dpdrmagnetic}-\eqref{dmdrmagnetic} boil down to the standard TOV equation (Eq.~\eqref{eq:2}).

We emphasize that the magnetic field in a neutron star can also affect the EoS, i.e., $\varepsilon$, and $P$. Due to Landau quantization magnetic field affects the dispersion relation of electrically charged particles~\cite{Broderick:2000pe,Strickland:2012vu}. Moreover, if we consider the spin magnetic field interaction, then the dispersion relation of the uncharged particles becomes modified~\cite{Broderick:2000pe,Strickland:2012vu}. These modified dispersion relations in the presence of a magnetic field also affect the nuclear matter EoS. However, in the present study, we do not consider the effect of the magnetic field on energy density ($\varepsilon$) and pressure ($P$). The rationale is as follows: leptons are not the dominant contributors to the equation of state, as they make only a small fraction of the matter inside the neutron star. Moreover, protons are electrically charged, but they are massive. The typical strength of the magnetic field that is required to affect the dynamics of a proton is of the order of $eB\sim M_{proton}^2\sim 1$ GeV$^2\sim 10^{20}$ Gauss, which is very large and five orders of magnitude larger than the surface magnetic field of a magnetar, which can be as large as $10^{15}$ Gauss. Of course, the magnetic field at the center can be even larger than the surface magnetic field. Therefore, the effect of the magnetic field on the equation of state is expected to be small. In the presence of a magnetic field, the pressure along the direction of the magnetic field $(P_{||})$ and perpendicular to the magnetic field ($P_{\perp}$) will be different~\cite{Patra:2020wjy}. In Ref.~\cite{Patra:2020wjy}, the authors argued that for a central magnetic field of the order of $10^{18}$ Gauss, the maximum mass corresponding to $P_{||}$, and  $P_{\perp}$ are very close. The quantitative asymmetry measure $\delta=(M_{\perp}-M_{||})/(M_{\perp}+M_{||})$ is only 0.5\%. Here $M_{\perp}$, and $M_{||}$ are the masses corresponding to $P_{\perp}$, and $P_{||}$, respectively. For an effective
chiral model EoS~\cite{Patra:2020wjy} in the presence of a magnetic field, maximum mass corresponding to $P_{\perp}$, and $P_{||}$ are 1.97 $M_{\odot}$ and 1.96 $M_{\odot}$, respectively. The radius of the neutron star corresponding to maximum mass is 11.43 km and 11.47 km, respectively. This implies that the pressure anisotropy can be small, leading to a very small asymmetry. Hence, we have not incorporated any effect of the magnetic field in the EoS, and have not used an anisotropic TOV framework. For completeness, the effect of the magnetic field on the EoS, and the properties of a dark matter admixed neutron star have been discussed in Ref.~\cite{Parmar:2023zlg,Routaray:2024fcq}. However, in these studies, the effect of the magnetic field on the TOV equation has not been accounted for. Hence, the present study can be considered as a complement to the studies done in Ref.~\cite{Parmar:2023zlg,Routaray:2024fcq}.

\subsection{Tidal Deformability}

We also compute the compactness and deformability of dark matter admixed neutron stars. The compactness of a star is defined as: 
\begin{align}
C = \frac{M}{R}.
\end{align}
Here $M$ is the mass, and $R$ is the radius of the star. In the neutron star binary systems, one companion's gravitational tidal field can induce a quadrupole moment in the other companion and can deform the shape of the star. Induced quadrupole moment is proportional to the tidal field of the companion in a binary via a proportionality constant called the tidal deformability. Tidal deformability can be obtained by solving the following equation, 
\begin{align}
r\frac{dy(r)}{dr} + (y(r))^2 + y(r) F(r) + r^2 Q(r) = 0,
\label{dydrequ}
\end{align}
with cofficients $F(r)$ and $Q(r)$ given as:
\begin{align}
& F(r) = \frac{r-4\pi r^3\bigg[\bigg(\varepsilon(r) + B^2/(8\pi)\bigg)-P(r)\bigg]}{r-2m(r)}\label{frmag}
\end{align}

\begin{widetext}
\begin{align}
& Q(r) = \frac{4 \pi r \bigg[\bigg(P(r)+\varepsilon+\frac{B^2}{8\pi}\Big)\frac{d\varepsilon}{dP} + 5\bigg(\varepsilon + \frac{B^2}{8\pi} \bigg) +9P -\frac{6}{4 \pi r^2} \Bigg]}{r-2m(r)} - 4 \bigg[\frac{4 \pi r^3 P(r) + m(r) }{r^2\bigg(1-\frac{2m(r)}{r}\bigg)}\bigg]^2 \label{qrmag}
\end{align}
\end{widetext}
along with the TOV equation (Eqs.~\eqref{dpdrmagnetic}-\eqref{dmdrmagnetic}). The detailed derivation of tidal deformability for a non-magnetized neutron star can be found in Refs.~\cite{Hinderer:2007mb,Das:2020ecp}. The crucial difference between the expressions of $F(r)$, and $Q(r)$ as given in Eqs.~\eqref{frmag}-\eqref{qrmag}, with  $F(r)$, and  $Q(r)$ for a non-magnetized neutron star, is the factor $B^2/(8\pi)$. This factor of $B^2/(8\pi)$ in  $F(r)$, and  $Q(r)$ originates from the $B^2/(8\pi)$ factor in the TOV equation \footnote{In the absence of a magnetic field the radial evolution of metric coefficient $\beta(r)$ can be expressed as $r \beta^{\prime} = 1-e^\beta + 8\pi r^2 \varepsilon e^\beta$~\cite{Das:2020ecp}. From the TOV equations (Eqs.~\eqref{dpdrmagnetic}-\eqref{dmdrmagnetic}), we observe that the energy density ($\varepsilon$) is replaced by $\varepsilon+B^2/(8\pi)$, in the presence of a magnetic field. This additional magnetic field dependent term also arises in the expression of $d\beta/dr$, i.e., $r \beta^{\prime} = 1-e^\beta + 8\pi r^2 (\varepsilon + \frac{B^2}{8\pi})e^\beta$ affecting the TOV equation and the tidal deformability equation.} Eqs.~\eqref{dpdrmagnetic}-\eqref{dmdrmagnetic}. The dimensionless tidal deformability, denoted by $\Lambda$, is expressed as:
\begin{align}
\Lambda = \frac{3k_2}{2C^5},
\end{align}
where $k_2$ is the tidal love number defined as, 
\begin{align}
& k_2 = \frac{8C^5}{5}(1-2C)^2[2+2C(y_R-1)-y_R] \nonumber\\
& \times \Big\{2C(6-3y_R+3C(5y_R-8)) \nonumber\\
& +4C^3[13-11y_R+C(3y_R-2)+2C^2(1+y_R)] \nonumber\\
& +3(1-2C)^2[2-y_R+2C(y_R-1)]\log(1-2C)\Big\}^{-1}.
\end{align}
The quantity $y_R\equiv y(R)$ is the solution of Eq.~\eqref{dydrequ} at $r=R$. The boundary condition to solve Eq.~\eqref{dydrequ} has been discussed in Ref.~\cite{Hinderer:2007mb}.

\begin{figure}[]
\includegraphics[width=0.5\textwidth]{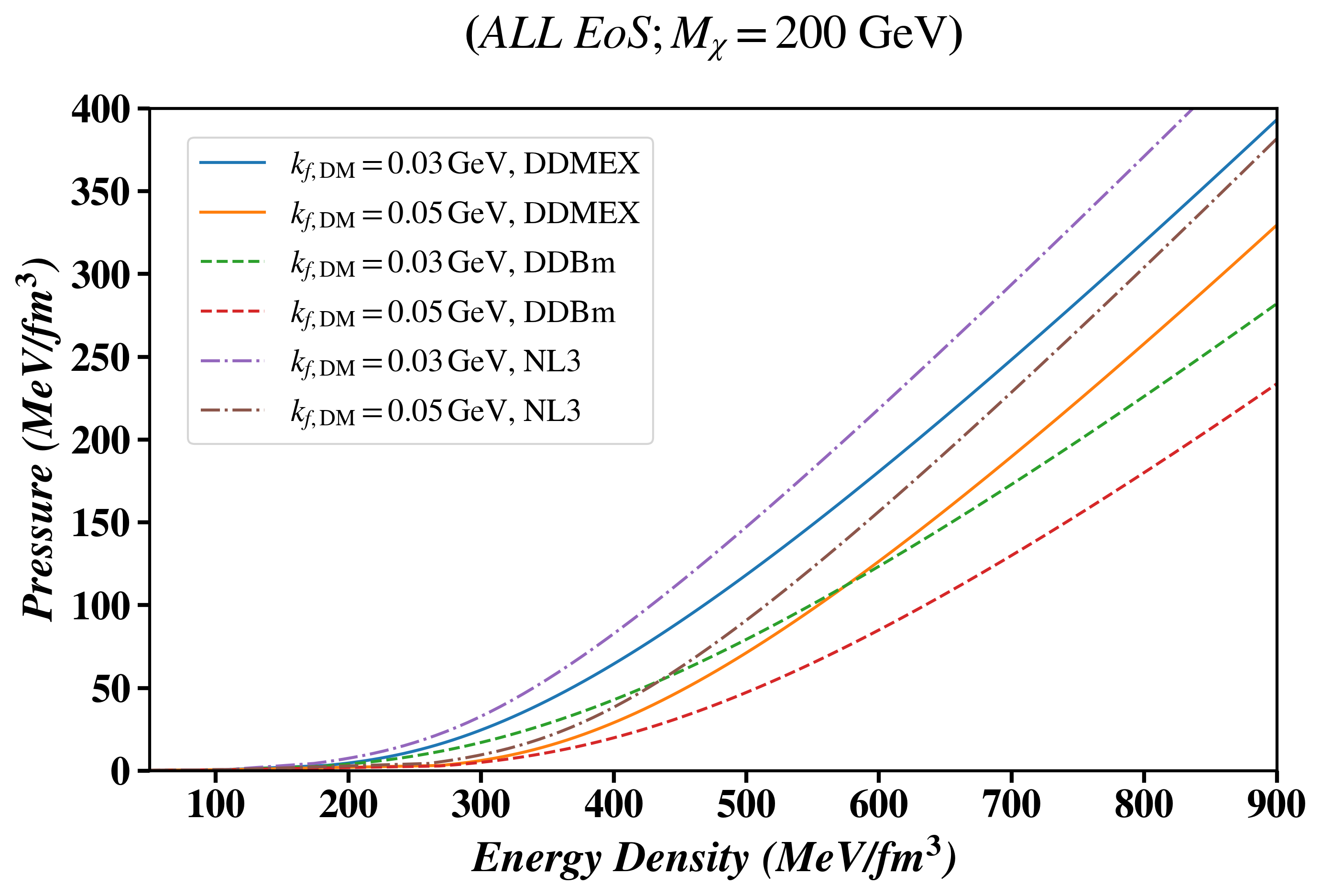} 
\caption{Variation of different EoS with the dark matter Fermi momentum. With an increase in the dark matter Fermi momentum, the EoS becomes softer. For this plot, we have used bare mass of the dark matter particle $M_{\chi}=200$ GeV.} 
\label{fig:ALL_EoS_fixmass}
\end{figure}

\begin{figure}[]
\centering
\includegraphics[width=0.5\textwidth]{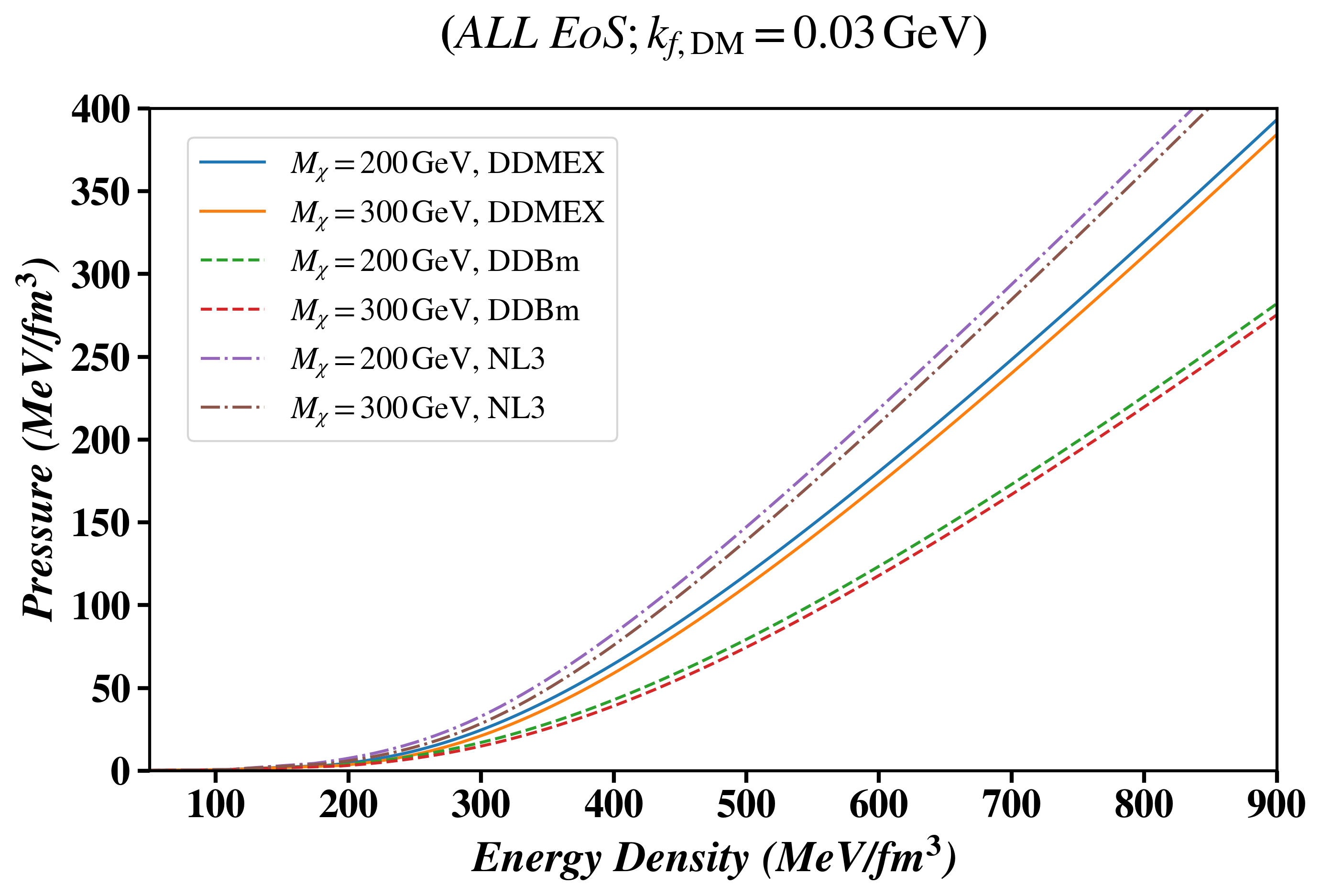} 
\caption{Variation of different EoS with the dark matter mass ($M_{\chi}$). With an increase in the dark matter mass the EoS becomes softer. For this plot we have used dark matter Fermi momentum $k_{f,DM}=0.03$ GeV.} 
\label{fig:ALL_EoS_fix_Kf1}
\end{figure}

\begin{figure}[]
\centering
\includegraphics[width=0.45\textwidth]{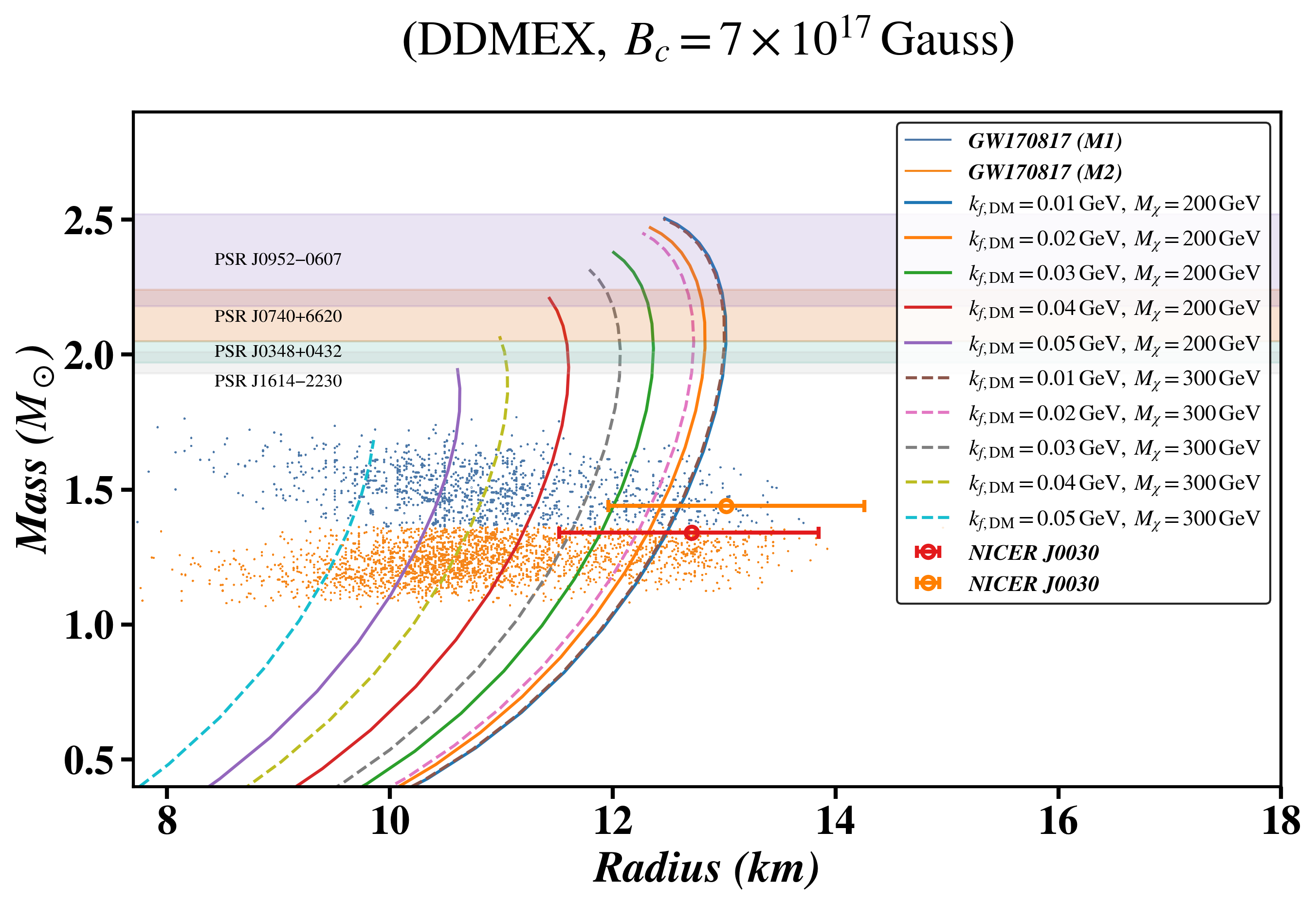}
\includegraphics[width=0.45\textwidth]{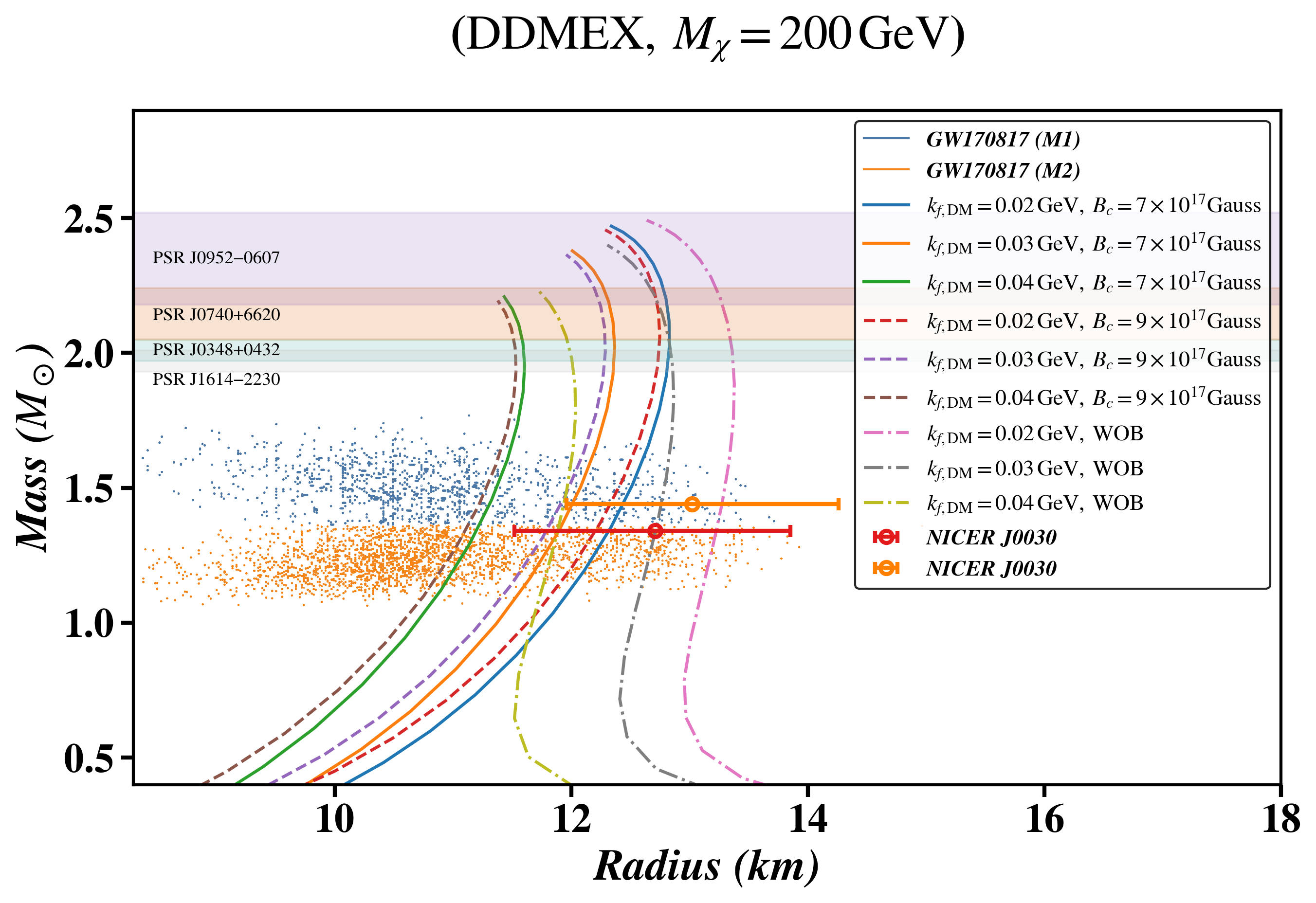}
\caption{Upper panel: $M-R$ sequences for DDMEX EoS. One observes the effect of changing dark matter Fermi momentum $k_{f,DM} = 0.01-0.05$ GeV and mass $M_{\chi} = 200$ and $300$ GeV for a fixed value of central magnetic field $B_c = 7\times10^{17} \mathrm{Gauss}$. Lower panel: $M-R$ sequences for DDMEX EoS. This plot is for dark matter mass $M_{\chi} = 200$ GeV, and depicts the effect on the $M-R$ relation by taking two values for central magnetic field, i.e., $B_c = 7\times10^{17} \mathrm{Gauss},\;9\times10^{17} \mathrm{Gauss}$. The results for the scenario without a magnetic field (WOB) for $k_{f, DM} = 0.01-0.05$ GeV have also been shown here.}
\label{fig:mr_ddmex}
\end{figure}

\begin{figure}[]
\centering
\includegraphics[width=0.45\textwidth]{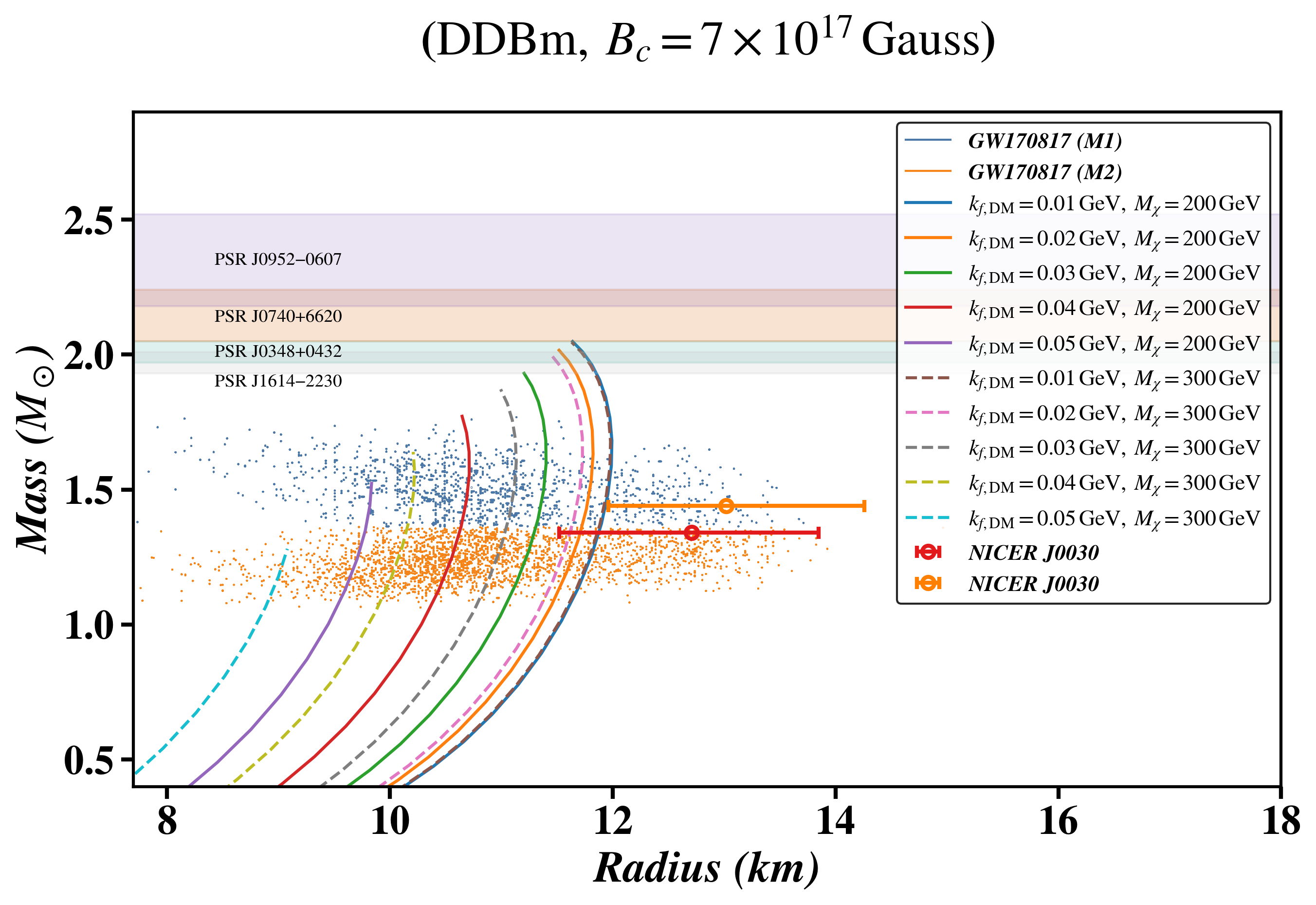}
\includegraphics[width=0.45\textwidth]{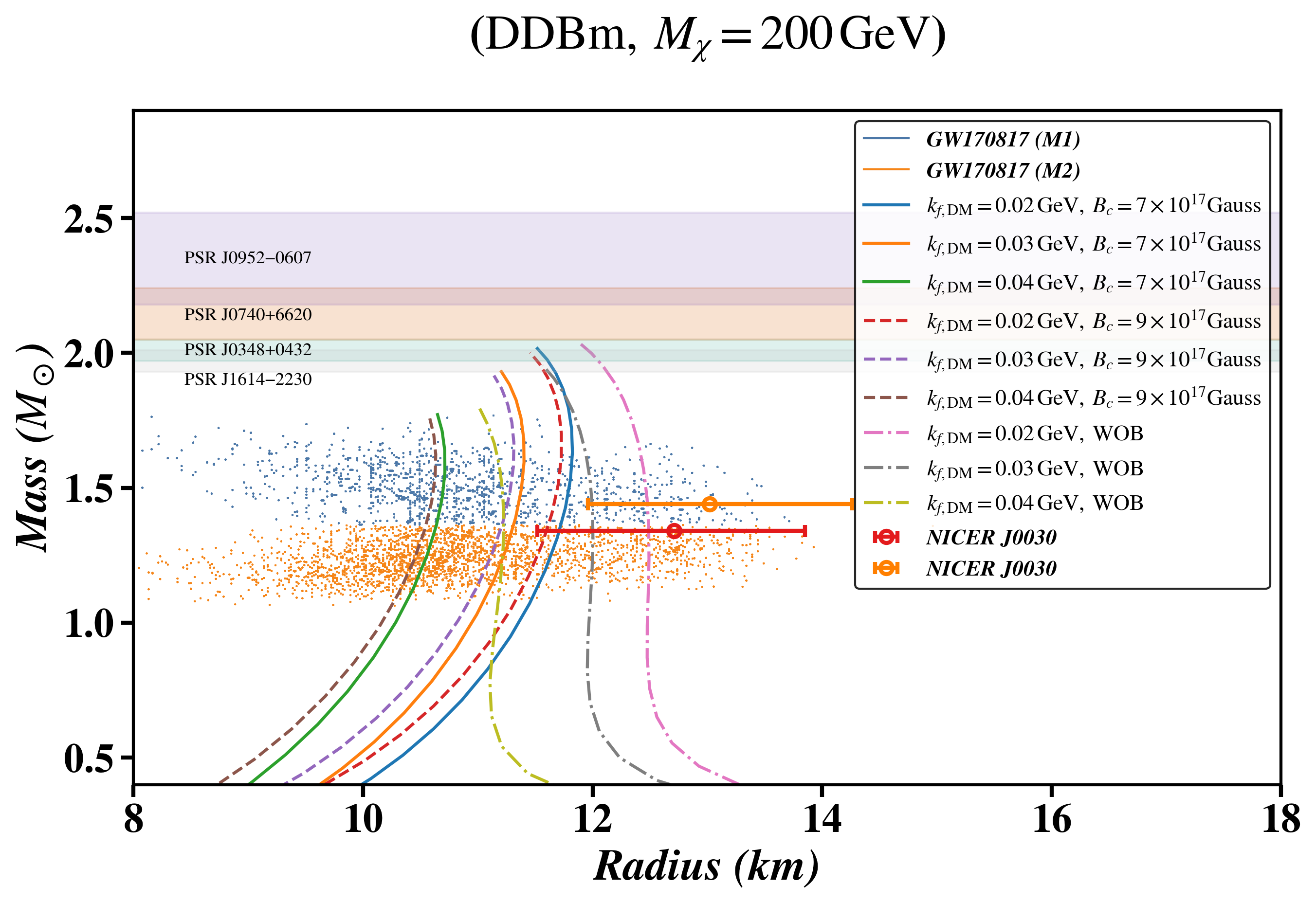}
\caption{$M-R$ sequences for DDBm EoS. In this figure, we showcase the $M-R$ relation as a function of dark matter mass ($M_{\chi}$), dark matter Fermi momentum ($k_{f,DM}$), and magnetic field ($B_c$). The range of $M_{\chi}$, $k_{f,DM}$, and $B_c$ are same as in Fig.~\eqref{fig:mr_ddmex}.}
\label{fig:mr_ddbm}
\end{figure}

\begin{figure}[]
\centering
\includegraphics[width=0.45\textwidth]{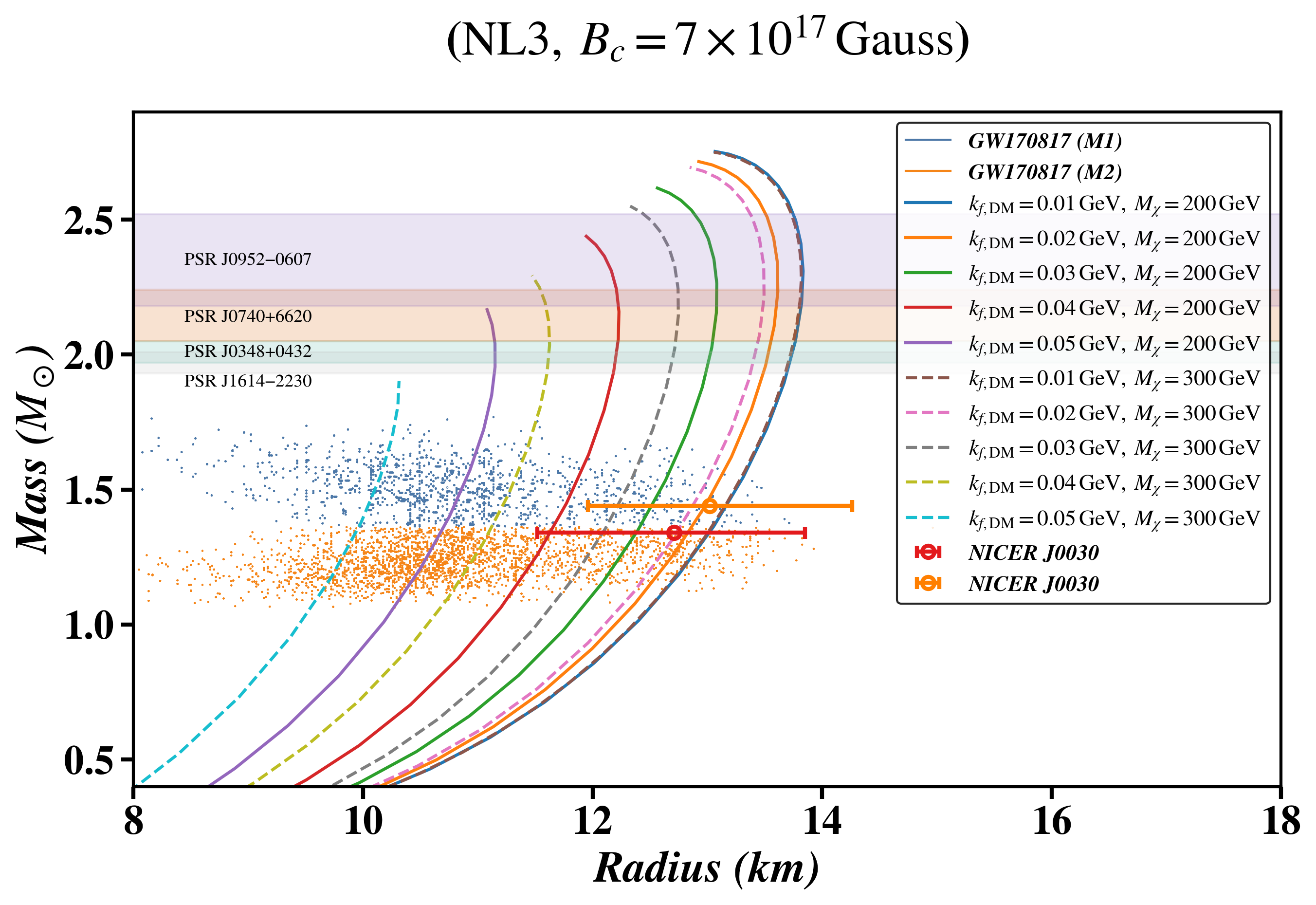}
\includegraphics[width=0.45\textwidth]{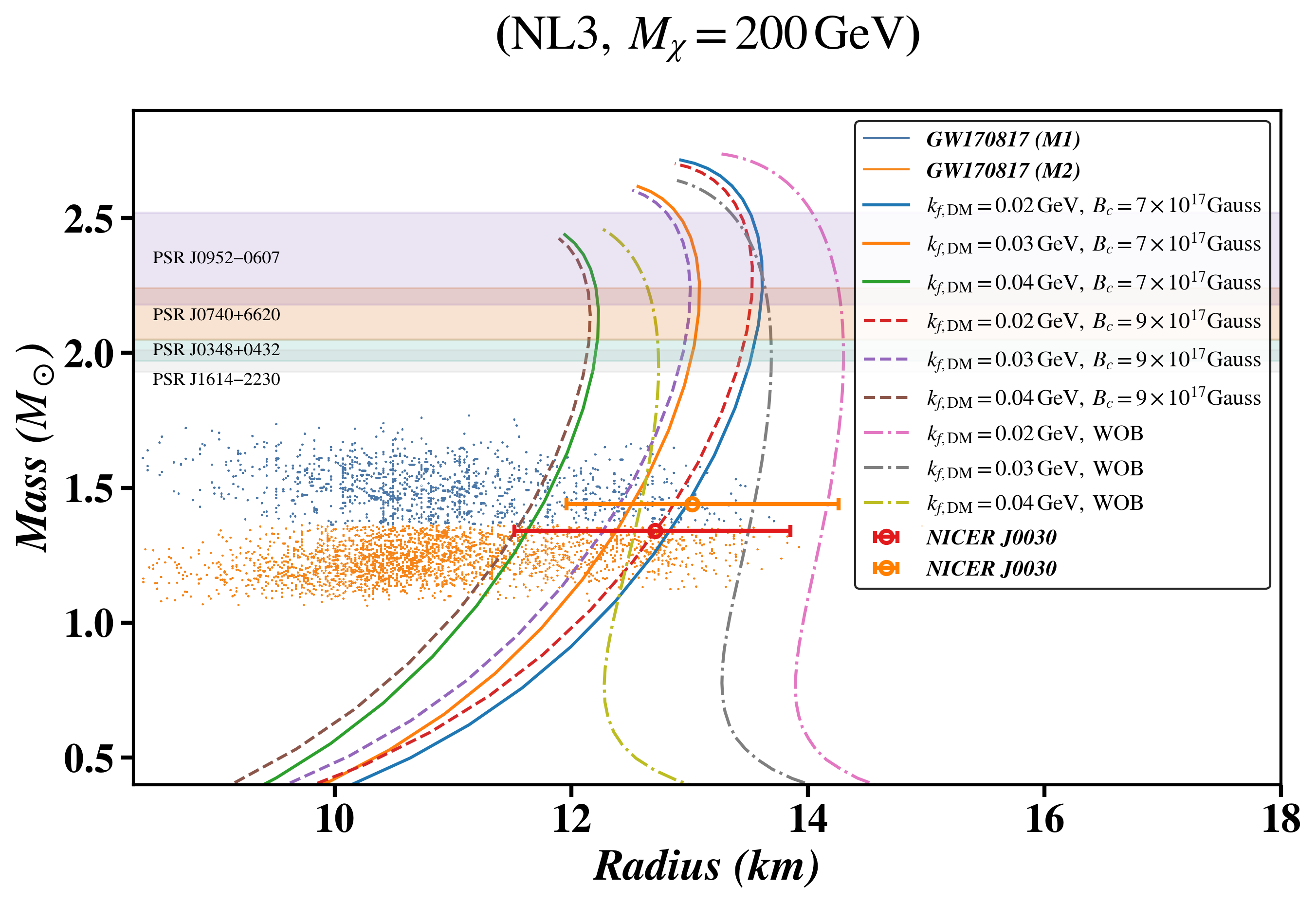}
\caption{$M-R$ sequences for NL3 EoS. In this figure, we showcase the $M-R$ relation as a function of dark matter mass ($M_{\chi}$), dark matter Fermi momentum ($k_{f,DM}$), and magnetic field ($B_c$). The range of $M_{\chi}$, $k_{f,DM}$, and $B_c$ are same as in Fig.~\eqref{fig:mr_ddmex}.}
\label{fig:mr_nl3}
\end{figure}

\begin{figure}[]
\includegraphics[width=0.5\textwidth,keepaspectratio]{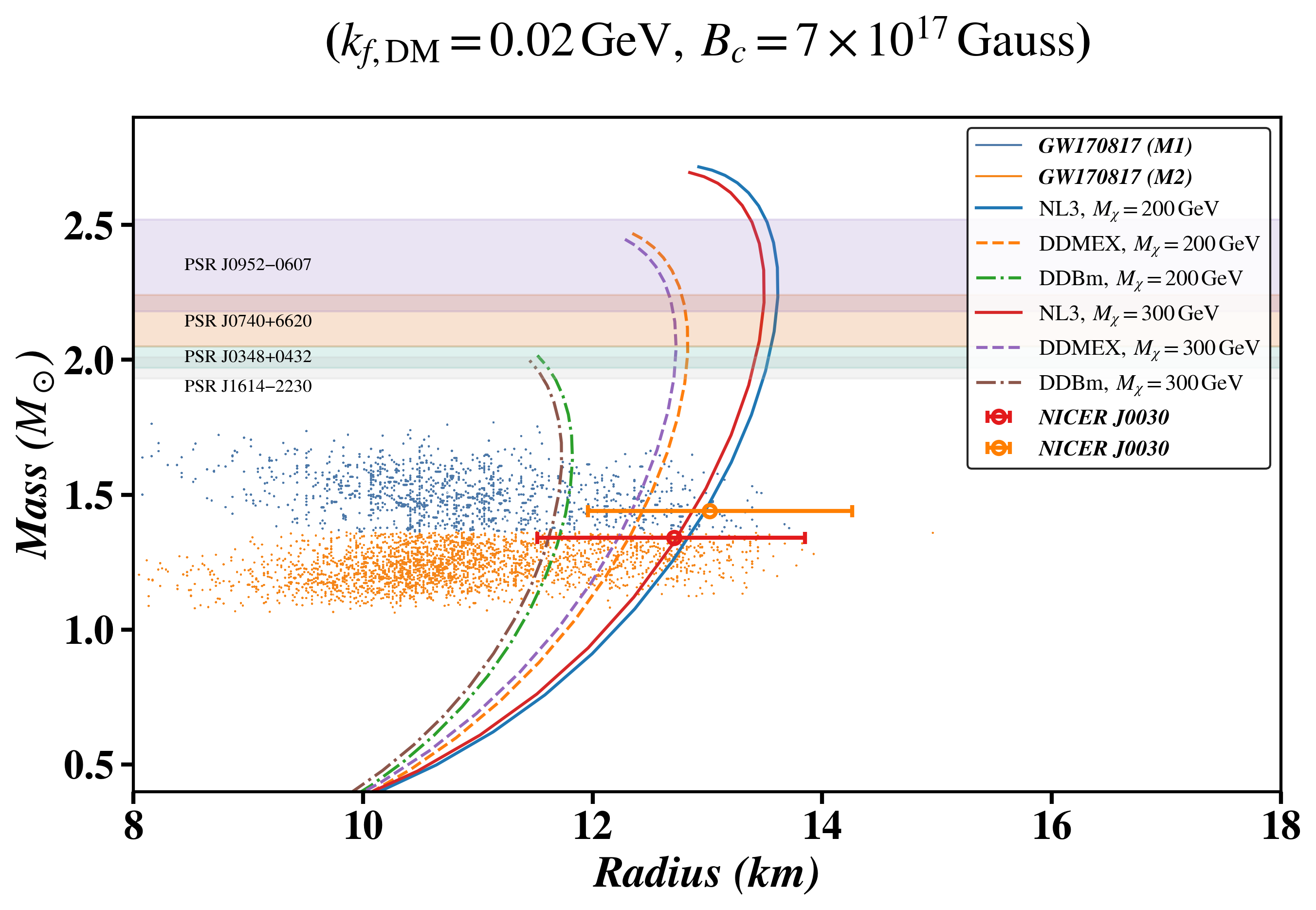} 
\caption{Comparison between $M-R$ sequences for NL3, DDMEX, DDBm EoSs for a fixed value of dark matter Fermi momentum $k_{f,DM} = 0.02$ GeV, and the central value of the magnetic field $B_c=7\times 10^{17}$ Gauss. We consider two values of dark matter mass, $M_{\chi}=200,~300$ GeV.}
\label{fig:all_eos_mr}
\end{figure}

\subsection{Non-radial oscillations}

In this section, we briefly discuss the non-radial oscillations of compact stars. For completeness, we present only the final set of coupled equations governing these oscillations. Their detailed derivation within the Cowling approximation can be found in refs.~\cite{Kumar:2024abb, Kumar:2021hzo, Das:2021dru} and the references therein. In the Cowling approximation, perturbations of the spacetime metric are neglected, while the fluid displacement is expanded in terms of spherical harmonics. The resulting coupled differential equations are then solved to determine the eigen frequencies of the non-radial fluid oscillation modes of neutron stars,
\begin{align}
& Q' - \frac{1}{c_s^2}\left[\omega^2 r^2 e^{\frac{\beta-2\alpha}{2}} Z + \frac{\alpha'}{2} Q \right]
+ l(l+1)e^\frac{\beta}{2} Z = 0,\\
& Z' - \alpha' Z + e^\frac{\beta}{2} \frac{Q}{r^2}
- \frac{\alpha}{2}'\left(\frac{1}{c_e^2} - \frac{1}{c_s^2}\right)\nonumber\\
&~~~~~~~~~~~~~~~~~~~~~~~~ \times
\left(
Z + \frac{\alpha'}{2} e^{\frac{-\beta+2\alpha}{2}}\frac{Q}{\omega^2 r^2}
\right)
= 0. \label{eq:26}
\end{align}
Here, a prime denotes a derivative with respect to the radial distance $r$ and $\omega$ is the eigen frequency of the non-radial oscillation of the fluid perturbation. These first-order coupled differential equations are solved with boundary conditions. The field perturbation functions $Q$ and $Z$ vary near the center of a star as following,
\begin{align}
Q(r) = A r^{l+1}, \qquad
Z(r) = -\, A \frac{r^l}{l},
\end{align}
where $A$ is an arbitrary constant. In the present work, we set $A=1$. We focus only on the quadrupolar ($l=2$) oscillation modes. The equation of state and these coupled differential equations are derived in units $G=c=\hbar=1$ and are therefore solved consistently in the same framework. In the present work, the last term in Eq. (\ref{eq:26}) is neglected. The coupled differential equations for $Q$ and $Z$ are integrated from center to the stellar surface for a given trial value of oscillation eigen frequencies. The correct solutions of the oscillation eigen frequencies are determined by iteratively adjusting the trial frequency until the surface boundary conditions given below are satisfied. The boundary condition is given below,
\begin{align}
\bigg(\omega^{2} r^{2} e^{\frac{\beta - 2\alpha}{2}} Z + \frac{\alpha'}{2} Q \bigg)_{r=R} = 0.
\end{align}

\section{Results and Discussions} \label{results_and_discussions}

\begin{figure*}
\centering
\includegraphics[width=0.8\textwidth,keepaspectratio]{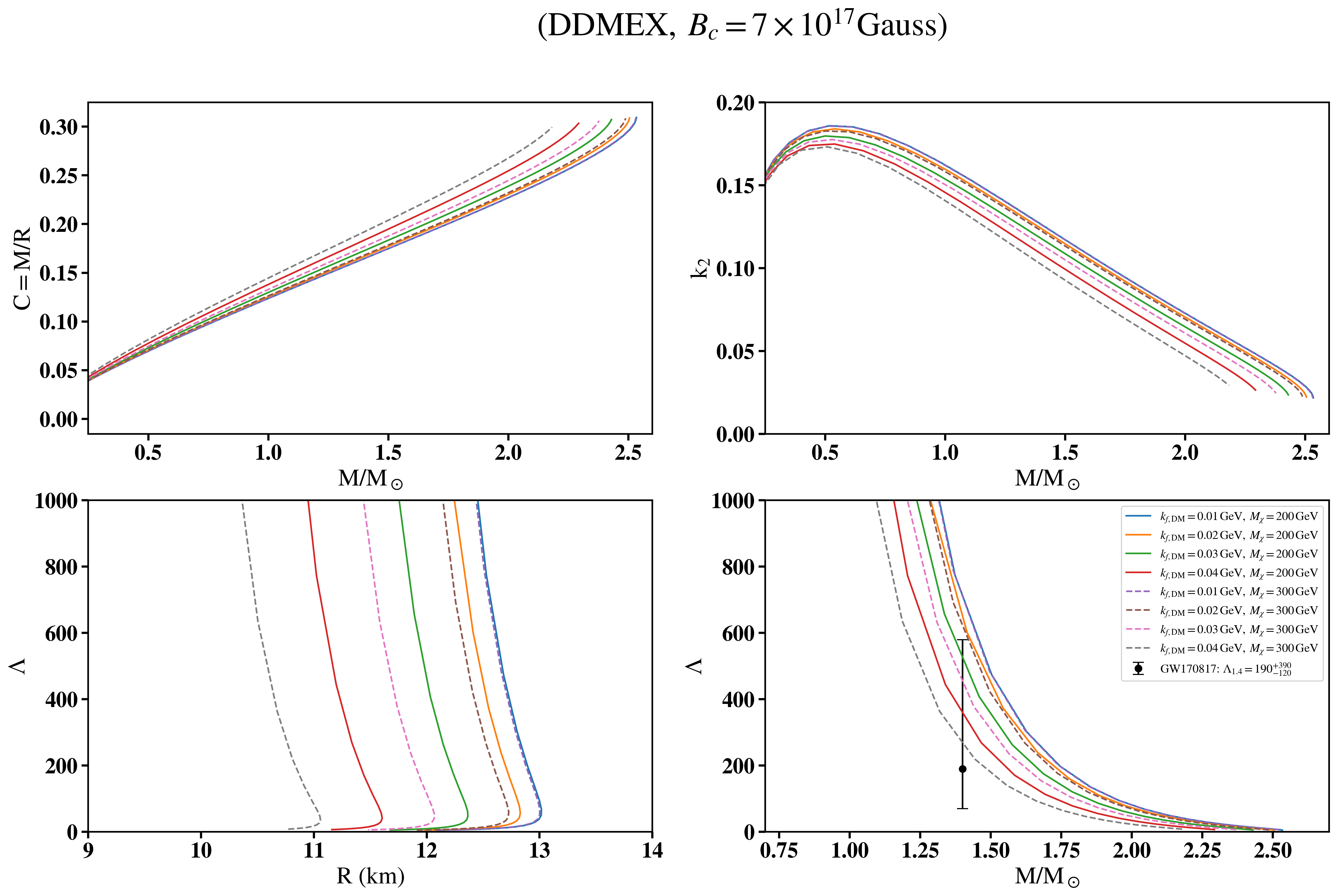} 
\caption{In the upper panel we showcase the variation of compactness $(M/R)$, tidal love number ($k_2$) with the neutron star mass ($M$). In the lower panel, we also show the variation of dimensionless tidal deformability ($\Lambda$) with the neutron star mass ($M$) and radius ($R$). Here we consider DDMEX EoS, $k_{f, DM}$ in the range $(0.01-0.04)$ GeV, $M_{\chi}$ in the range ($200, 300$) GeV. In this plot, the central value of the magnetic field is $B_c = 7\times10^{17}$ Gauss.}
\label{fig:DDMMEX_Bcfix_tidal}
\end{figure*}

\begin{figure*}
\centering
\includegraphics[width=0.8\textwidth,keepaspectratio]{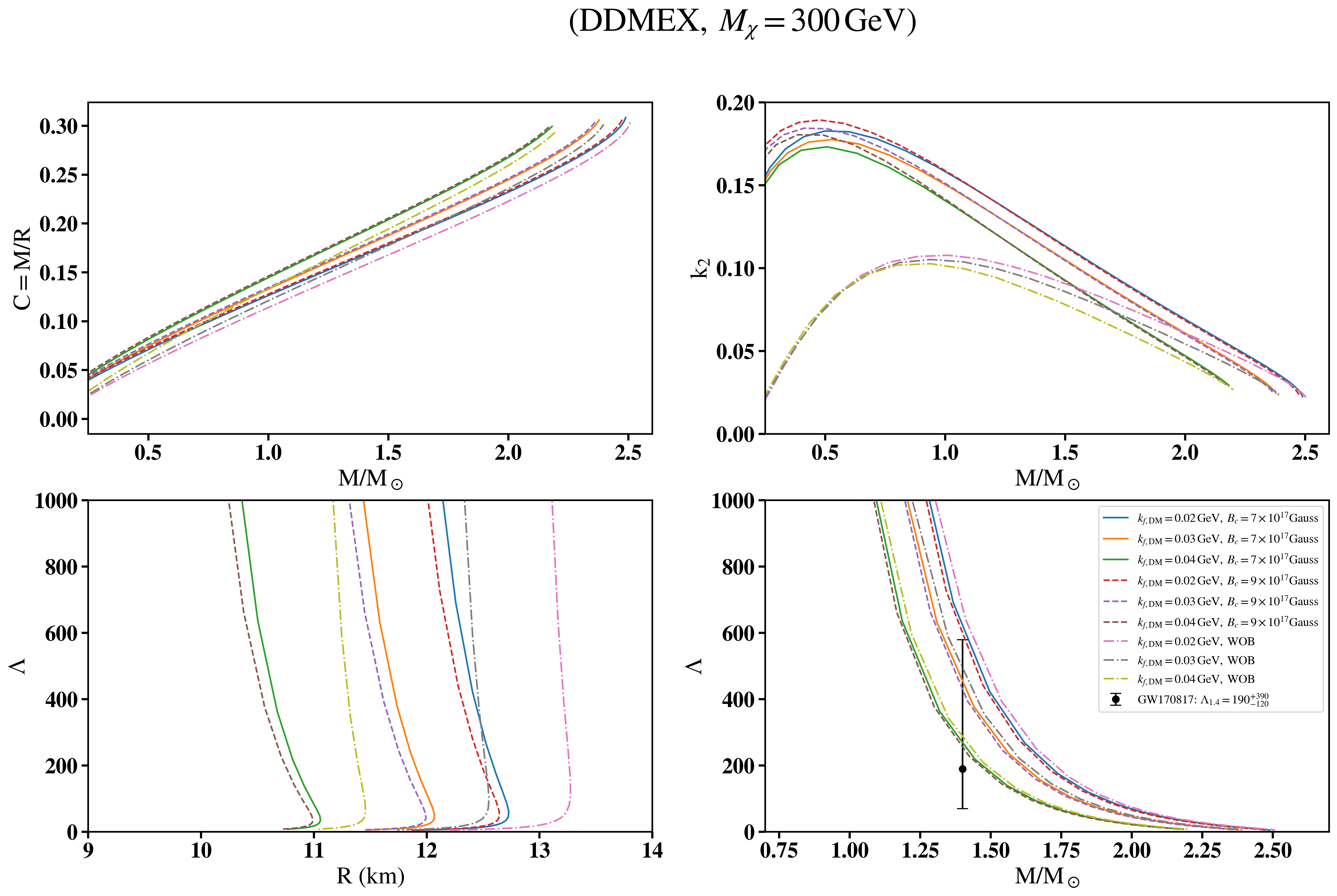} 
\caption{In the upper panel we showcase the variation of compactness $(M/R)$, tidal love number ($k_2$) with the neutron star mass ($M$). In the lower panel, we also show the variation of dimensionless tidal deformability ($\Lambda$) with the neutron star mass ($M$) and radius ($R$). Here, we consider DDMEX EoS. $k_{f, DM} = 0.02, 0.03, 0.04$ GeV and fixed dark matter mass $M_{\chi} = 300$ GeV. We compare the results for different values of the central magnetic field $B_c = 7\times10^{17}$ Gauss, $9\times10^{17}$ Gauss. Properties for a neutron star without a magnetic field ($WOB$) are plotted for comparison.}
\label{fig:DDMEX_Bcvar_tidal}
\end{figure*}

\begin{figure*}
\centering
\includegraphics[width=0.8\textwidth,keepaspectratio]{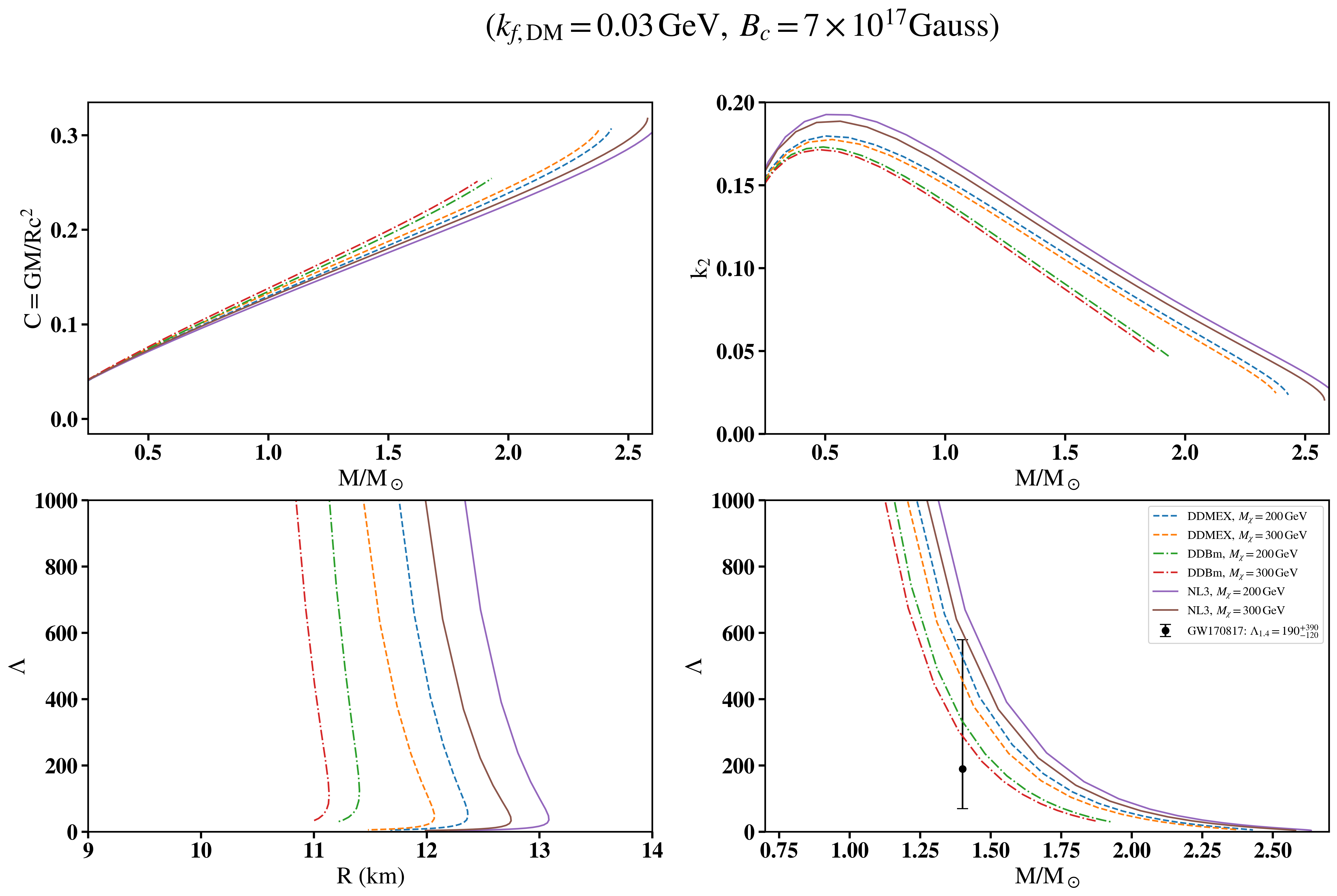} 
\caption{In the upper panel we compare the variation of compactness $(M/R)$, tidal love number ($k_2$) with the neutron star mass ($M$) for different EoSs. In the lower panel, we compare the variation of dimensionless tidal deformability ($\Lambda$) with the neutron star mass ($M$) and radius ($R$), for DDMEX, DDBm and NL3 EoSs. Here, the dark matter mass is taken as $M_{\chi} = 200, 300$ GeV for a fixed value of dark matter fermi momentum $k_{f, DM} = 0.03$ GeV and $B_c$ is taken equal to $7\times10^{17}$ Gauss.}
\label{fig:ALL_EOS_tidal}
\end{figure*}

In this section, we show the EoS, mass-radius, compactness, tidal love number, tidal deformability and non-radial oscillation frequency results for both density-dependent and density-independent EoS parameters. EoSs for density-dependent as well as density-independent scenarios are shown in Figs.~\eqref{fig:ALL_EoS_fixmass}, and \eqref{fig:ALL_EoS_fix_Kf1} for different values of dark matter fermi momenta, and dark matter masses, respectively. It is evident from Fig.~\eqref{fig:ALL_EoS_fixmass} that as we increase the value of the Fermi momentum, the dark matter energy density versus pressure curve shifts towards a lower value of pressure, i.e., the equation of state becomes softer. Also, Fig.~\eqref{fig:ALL_EoS_fix_Kf1} depicts EoS curves shifts towards lower pressure values with an increase in the mass of dark matter particle. These results are in accordance with the previously reported results, e.g., in Refs.~\cite{Das:2018frc,Panotopoulos:2017idn}. From these plots, we also observe that, among the three EoS used in this paper, DDMEX~\cite{Huang:2020cab} is softer than NL3~\cite{Das:2020ecp} and stiffer than DDBm~\cite{Malik:2022zol}.

In Figs.~\eqref{fig:mr_ddmex}-\eqref{fig:mr_nl3} we show the mass-radius sequences of neutron stars for DDMEX, DDBm and NL3 equations of state, respectively. In these plots, we vary the Fermi momentum of dark matter, the mass of the dark matter particle, and the central value of the magnetic field. In Figs.~\eqref{fig:mr_ddmex}-\eqref{fig:mr_nl3}, the upper panel represents the effect of dark matter Fermi momentum and dark matter mass on $M-R$ relation for a fixed value of central magnetic field, $B_c = 7\times10^{17} \mathrm{Gauss}$. We emphasize that the equation of states do not depend on the magnetic field, but the magnetic field dependence of the $M-R$ plot arises from the modified TOV equations~\eqref{dpdrmagnetic}-\eqref{dmdrmagnetic}.  In the upper panel of these plots (Figs.~\eqref{fig:mr_ddmex}-\eqref{fig:mr_nl3}), for each EoS, the solid lines show the results for $M_{\chi} = 200 \mathrm{GeV}$ and the dashed curve represents the same with $M_{\chi} = 300 \mathrm{GeV}$ case. It is clearly visible from plots that for all three EoSs, the $M-R$ curve shifts towards a lower value of maximum mass and a lower value of radius with an increase in either dark matter Fermi momentum or dark matter mass. In the lower panels in all three Figs.~\eqref{fig:mr_ddmex}-\eqref{fig:mr_nl3} we show the effect of magnetic field on the mass- radius sequences of a dark matter admixed neutron star. In the lower panels of these plots, we show the $M-R$ plots for a fixed value of the dark matter mass $M_{\chi} = 200~\mathrm{GeV}$, but vary the central magnetic field and the Fermi momentum of dark matter. We consider two values of the central magnetic field, $B_c = 7\times10^{17} \mathrm{Gauss}$, and $9\times10^{17} \mathrm{Gauss}$. Results for non-magnetized stars are also included in the same plots for comparison. From the lower panels, one observes that the $M-R$ curve shifts towards lower maximum mass and lower radius, with an increase in $B_c$. 

Some comparison with already existing results on the effect of the magnetic field on $M-R$ plots is in order here. In Ref.~\cite{Parmar:2023zlg}, the authors study the effect of the magnetic field on the dark matter admixed neutron stars. However, they have considered the effect of the magnetic field in the EoS, but not in the TOV equation. In Ref.~\cite{Parmar:2023zlg} authors argue that, in the presence of a magnetic field, the EoS has an anisotropic structure; the pressure along the direction of the magnetic field ($P_{||}$) becomes softer with magnetic field, and the pressure perpendicular to the direction of the magnetic field ($P_{\perp}$) becomes stiffer with magnetic field~\cite{Parmar:2023zlg}. Such behaviour of the equation of state has also been reported for a magnetized neutron star in the absence of any dark matter component~\cite{Patra:2020wjy}. This also affects the $M-R$ relation. The $M-R$ relation calculated with $P_{||}$ indicates that with an increase in the magnetic field, the maximum mass and the corresponding radius decrease. This is similar to our results. But when they consider the $M-R$ relation with $P_{\perp}$, the maximum mass and the corresponding radius increase with magnetic field. This is the opposite of our findings.

In Fig.~\eqref{fig:all_eos_mr} we compare the $M-R$ sequences for different EoSs by varying the mass of dark matter particle. Here we consider $M_{\chi} = 200$ and  $300$ GeV. But we have fixed the dark matter Fermi momentum $k_{f,DM} = 0.02$ GeV, and the magnetic field $B_c = 7\times10^{17} \mathrm{Gauss}$. Solid, dashed and dot-dashed curves represent $M-R$ relation for NL3, DDMEX, DDBm EoSs,  respectively. Sequences for EoS with density-dependent couplings lie towards lower mass and radius compared to EoS with density-independent couplings. Also, DDBm curves lie below DDMEX curves. In this $M-R$ plot we compare our results with the observational data for mass and radius for pulsars PSRJ0952-0607~\cite{Romani:2022jhd}, PSRJ0740+6620 ~\cite{Fonseca:2021wxt}, PSRJ0348+0432 ~\cite{Antoniadis:2013pzd}, PSRJ1614-2230~\cite{Demorest:2010bx}; data sets NICERJ0030 ~\cite{Riley:2019yda} from neutron star's interior composition explorer and gravitational wave data GW170817(M1)~\cite{LIGOScientific:2018cki}, GW170817(M2)~\cite{LIGOScientific:2018cki}. Here, $M1$ and $M2$ represent two companions in a binary inspiraling system and are also plotted to compare with the predicted sequences. The results obtained for all EoSs, as depicted in Figs.~\eqref{fig:mr_ddmex}-\eqref{fig:mr_nl3}, and \eqref{fig:all_eos_mr} follow the gravitational observations as plotted GW contours and the NICER observation as plotted vertical lines. Hence, it can be concluded that our predictions are consistent with the observation data points.

We show the compactness and tidal deformation properties in Figs.~\eqref{fig:DDMMEX_Bcfix_tidal}, \eqref{fig:DDMEX_Bcvar_tidal}, and \eqref{fig:ALL_EOS_tidal}. Each figure has four sub-figures. The upper panels in these figures show the results for compactness ($C$) and tidal love number ($k_2$), respectively. Both are plotted against the mass ($M$) of neutron stars. We represent the mass of the neutron star in units of solar mass ($M_{\odot}$). The lower panel in Figs.~\eqref{fig:DDMMEX_Bcfix_tidal}-\eqref{fig:ALL_EOS_tidal} we show the variation of tidal deformability ($\Lambda$) with the radius and mass (again in solar mass units), respectively.

In Fig.~\eqref{fig:DDMMEX_Bcfix_tidal}
we show the results for compactness ($C$), tidal love number ($k_2$), and tidal deformability ($\Lambda$) for DDMEX equation of state. Here Fermi momentum of dark matter is varied from $0.01 - 0.04$ GeV, and two values of $M_{\chi} = 200, 300$ GeV for the dark matter mass. In this figure, the value of the central magnetic field is $B_c = 7 \times 10^{17}$ Gauss. We observe that with an increase in $k_{f,DM}$ and $M_{\chi}$, compactness increases, but the tidal love number curve shifts towards a lower value. At the same time, tidal deformability curves shift towards a lower value of radius and a lower value of maximum mass. 

Fig.~\eqref{fig:DDMEX_Bcvar_tidal} explores the effect of changing the central magnetic field on compactness ($C$), tidal love number ($k_2$), and tidal deformability ($\Lambda$) for DDMEX equation of state. For comparison, we have also shown these properties for a  
non-magnetized neutron star. In this plot, we consider a fixed dark matter mass $M_{\chi}=300$ GeV, and a range of dark matter Fermi momenta $k_{f, DM} = 0.02, 0.03, 0.04$ GeV. We also consider two values of magnetic field, i.e., $B_c=7\times 10^{17}$ Gauss, and $B_c=9\times 10^{17}$ Gauss. We observe that the compactness increases with an increase in the value of the central magnetic field. For comparison, we also show the results for the zero magnetic field case (WOB). Curves representing the compactness in the presence of the magnetic field lie above the curve without the magnetic field. Moreover, the compactness also increases with $k_{f, DM}$. We find that the tidal love number increases with $k_{f,DM}$ and $B_c$. Tidal deformability curves shift towards lower mass and lower radius with increasing $k_{f,DM}$ and increasing $B_c$. For other EoSs, compactness ($C$), tidal love number ($k_2$), and tidal deformability ($\Lambda$) show similar behaviour (not shown here explicitly) with the dark matter mass ($M_{\chi}$), Fermi momentum ($k_{f,DM}$), and magnetic field ($B_c$),  as we find for DDMEX EoS.

In Fig.~\eqref{fig:ALL_EOS_tidal}, we compare 
compactness ($C$), tidal love number ($k_2$), and tidal deformability ($\Lambda$) for various EoSs for different values of the dark matter mass, $M_{\chi}=200, 300$ GeV. In this plot $k_{f,DM}$ and $B_c$ are fixed to $0.03$ GeV and $7\times10^{17}$ Gauss, respectively. We find that for the given set of parameters, the compactness is lowest for the NL3 EoS, and highest for the DDBm EoS. On the other hand, the tidal love number ($k_2$) is lowest for the DDBm EoS, and highest for the NL3 EoS. In the tidal deformability plot, curves for DDMEX lie below to NL3 and above to DDBm. Here, below (above) means towards the region of lower (higher) mass and lower (higher) radius.

\begin{figure}[]
\centering
\includegraphics[width=0.48\textwidth,keepaspectratio]{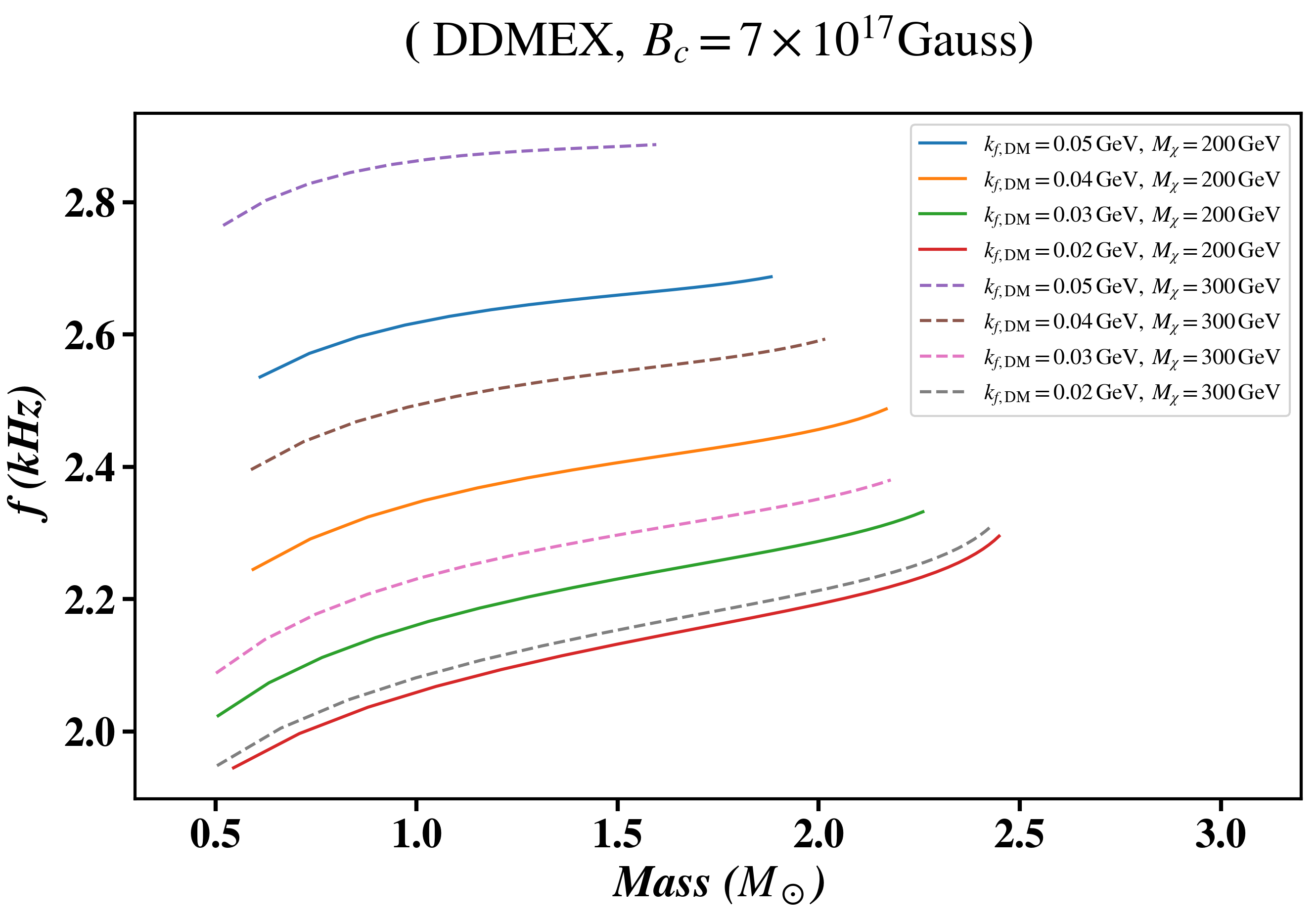}
\caption{Non-radial oscillation frequencies for DDMEX EoS. Here, the central value of the magnetic field is taken as $B_c = 7\times10^{17}$ Gauss, and $M_{\chi} = 200, 300$ GeV. In this plot, the dark matter Fermi momentum ($k_{f, DM}$) is taken in the range (0.02-0.05) GeV.}
\label{fig:DDMEX_BCfix_freq}
\end{figure}

\begin{figure}[]
\centering
\includegraphics[width=0.48\textwidth,keepaspectratio]{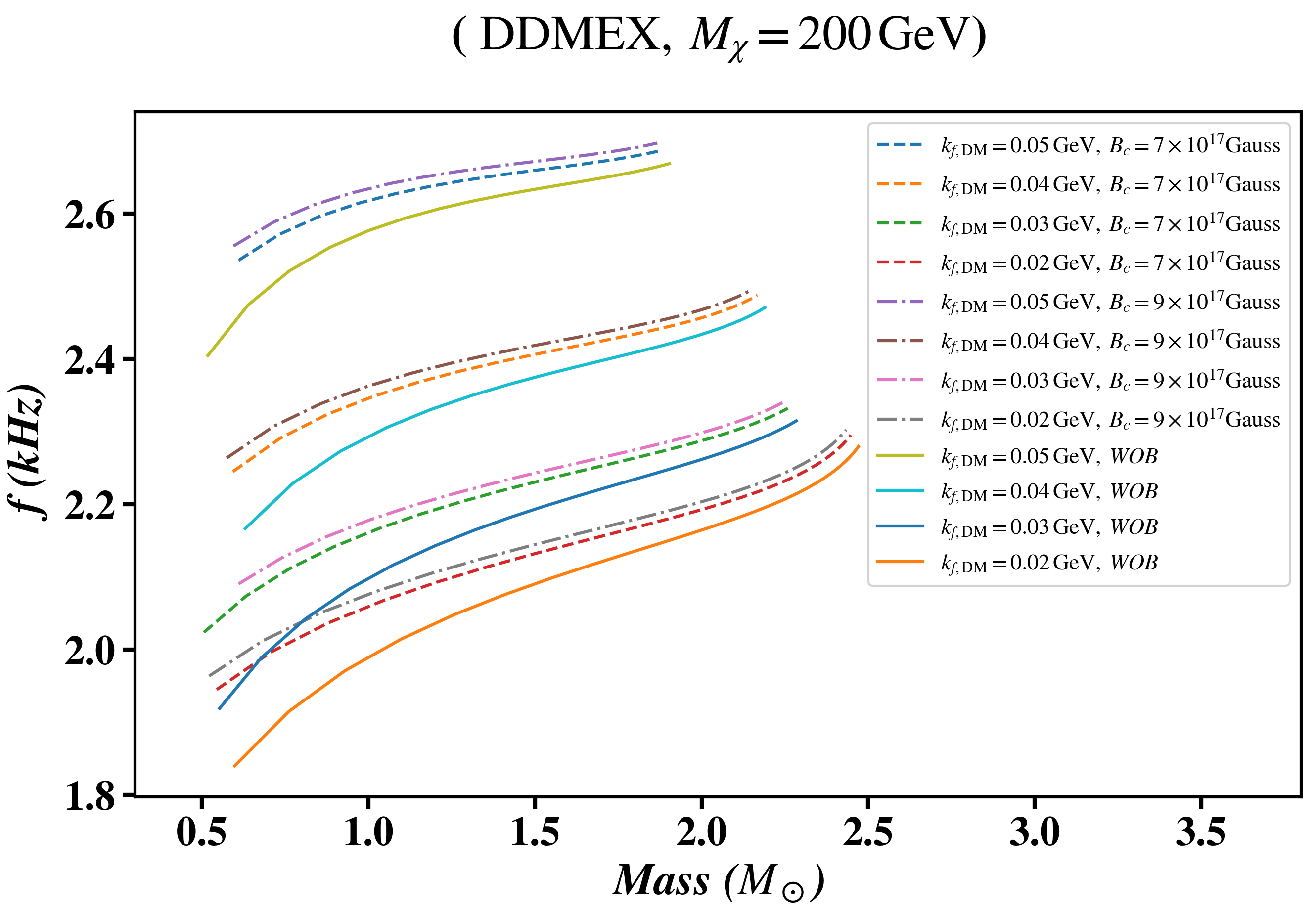}
\caption{Non-radial oscillation frequencies for DDMEX EoS. Here we consider two different values of the central magnetic field $B_c = 7\times10^{17}$ Gauss and $B_c = 9\times10^{17}$ Gauss. The dark matter mass is taken as $M_{\chi} = 200$ GeV. In this plot, the dark matter Fermi momentum ($k_{f, DM}$) is taken in the range (0.02-0.05) GeV. We also show the zero magnetic field results.}
\label{fig:DDMEX_BCvar_freq}
\end{figure}

\begin{figure}[]
\centering
\includegraphics[width=0.48\textwidth,keepaspectratio]{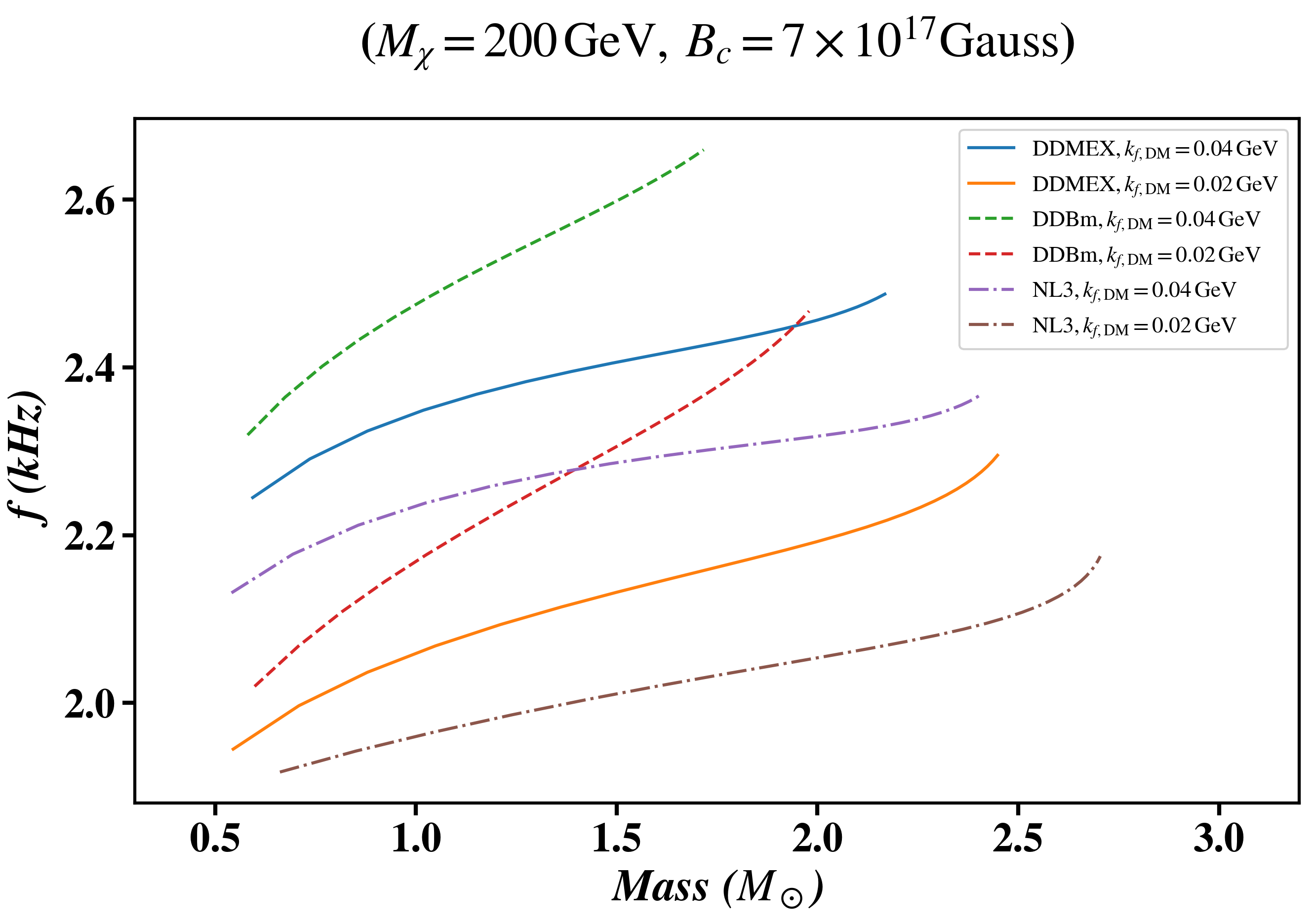}
\caption{Comparison of non-radial oscillation frequencies for DDMEX, DDBm, NL3 EoSs. We consider fixed values of dark matter mass $M_{\chi}=200$ GeV, and magnetic field $B_c=7\times 10^{17}$ Gauss. We consider dark matter Fermi momentum $k_{f,DM}=0.02-0.04$ GeV, for each EoSs.}
\label{fig:ALLEOS_BCfix_freq}
\end{figure}

Finally, in Figs.~\eqref{fig:DDMEX_BCfix_freq}, \eqref{fig:DDMEX_BCvar_freq}, and \eqref{fig:ALLEOS_BCfix_freq}, we show the effect of different EoSs, magnetic field, and dark matter properties ($M_{\chi}$, and $k_{f, DM}$) on the non-radial oscillation frequencies. In Fig.~\eqref{fig:DDMEX_BCfix_freq}, we show the variation of the non-radial oscillation frequencies with the neutron star mass (in units of solar mass) for different values of dark matter Fermi momentum and dark matter particle mass, for a specific value of the magnetic field. In Fig.~\eqref{fig:DDMEX_BCfix_freq}, we have considered DDMEX EoS. In Fig.~\eqref{fig:DDMEX_BCvar_freq} also we considered 
DDMEX EoS, but here show the variation of non-radial oscillation frequencies with the magnetic field and dark matter Fermi momentum. Finally, in Fig.~\eqref{fig:ALLEOS_BCfix_freq} we compare the results for different EoSs. In Fig.~\eqref{fig:ALLEOS_BCfix_freq}, we compare the variation of non-radial oscillation modes versus dark matter mass for three different EoS for a specific value of the dark matter mass and magnetic field, but different values of dark matter Fermi momentum.

\section{Conclusion and Future Outlook}
\label{conclusion}
In this investigation, we have explored the $M-R$ relation, compactness, and tidal properties of dark matter admixed magnetized neutron stars. In this study, we assume a non-vanishing fermionic dark matter component within the neutron star. The dark matter component interacts with the normal matter component through the Higgs portal interaction. We have considered Fermionic dark matter uniformly distributed inside the neutron stars, i.e., they have a fixed Fermi momentum. We have used the relativistic mean field approach to construct the EoS of dark matter admixed neutron stars. For a realistic modelling of the normal matter component, we use a density-independent and a density-dependent equation of state.  We also consider a radially dependent internal magnetic field to study its effect on the properties of dark-matter-admixed neutron stars.
In the present calculation, we do not consider the effect of the magnetic field on the neutron star EoS, but we incorporate the same through a modified TOV equation. This qualitatively modified TOV equation can be used to study the effect of the magnetic field on the neutron star properties without going into the detailed microscopic calculation. We take two different values of the magnetic field at the center of the neutron star, $B_c = 7\times10^{17}, 9\times10^{17}$ Gauss. Moreover, we consider EoS at zero temperature.

Because we are considering a fermionic dark matter particle with a mass larger than the nuclear mass, we find softening of the neutron star EoS with the mass of the dark matter and density. Since we are considering a uniformly distributed dark matter component, dark matter Fermi momentum carries the information of the dark matter density. To estimate the EoS of the dark matter admixed neutron star, we consider the dark matter Fermi momentum in the range $k_{f, DM} = 0.01-0.06$ GeV, dark matter mass in the range $M_{\chi} = 200, 300$ GeV. Using the neutron star EoS and the magnetic field-dependent TOV equation, we estimate the $M-R$ relation, compactness ($C$), tidal love number ($k_2$), and dimensionless tidal deformability. 
We find that with an increase in the dark matter mass, density (Fermi momentum), and magnetic field strength, the maximum mass and radius of a neutron star decrease. 
This also affects the compactness, tidal love number,  tidal deformability, and non-radial oscillation mode. The tidal deformability decreases with increasing dark matter mass, density (Fermi momentum), and magnetic field strength. On the other hand, the compactness and frequency of non-radial oscillations increase. We also compare these results with the available observational data. All computed results show reasonably good agreement with pulsar data sets and GW data for the full range of parameters used in this work. There are certain directions to improve or generalize the current calculation. One obvious direction is the consistent treatment of the magnetic field and system anisotropy on the EoS and the TOV equation. Also, we only consider zero-temperature EoS, but proto-neutron stars can have finite temperature~\cite{Issifu:2025jac}. Hence, a finite temperature generalization of EoS would be relevant to study its effects on a dark matter admixed magnetized neutron star.

\section*{Acknowledgments}
The authors thank the Department of Physics, BITS Pilani, Pilani Campus, for providing the necessary facilities and administrative support required for doing the work. One of the authors, Neeshu Rani, acknowledges CSIR New Delhi for providing financial support in terms of Junior Research Fellowship - CSIR. A.D. acknowledges the Anusandhan National Research Foundation (ANRF), Advanced Research Grant (ARG), project number: ANRF/ARG/2025/000691/PS. \\


\bibliography{ref.bib}

@article{Kumar:2024abb,
    author = "Kumar, Deepak and Mishra, Hiranmaya",
    title = "{CP violation in cold dense quark matter and axion effects on the non-radial oscillations of neutron stars}",
    eprint = "2411.17828",
    archivePrefix = "arXiv",
    primaryClass = "hep-ph",
    doi = "10.1088/1475-7516/2026/05/004",
    journal = "JCAP",
    volume = "05",
    pages = "004",
    year = "2026"
}

@article{Kumar:2026hoq,
    author ="Kumar, Deepak and Mohapatra, Ranjita K. and Mishra, Hiranmaya and Patra, Sudhanwa",
    title = "{Neutron star with dark matter using vector portal}",
    eprint = "2604.04560",
    archivePrefix = "arXiv",
    primaryClass = "hep-ph",
    doi = "10.48550/arXiv.2604.04560",
    journal = "",
    volume = " ",
    pages = " ",
    year = " "
}

@article{Kumar:2021hzo,
    author = "Kumar, Deepak and Mishra, Hiranmaya and Malik, Tuhin",
    title = "{Non-radial oscillation modes in hybrid stars: consequences of a mixed phase}",
    eprint = "2110.00324",
    archivePrefix = "arXiv",
    primaryClass = "hep-ph",
    doi = "10.1088/1475-7516/2023/02/015",
    journal = "JCAP",
    volume = "02",
    pages = "015",
    year = "2023"
}

@article{Leinson:2021ety,
    author = "Leinson, Lev B.",
    title = "{Impact of axions on the Cassiopea A neutron star cooling}",
    eprint = "2105.14745",
    archivePrefix = "arXiv",
    primaryClass = "hep-ph",
    doi = "10.1088/1475-7516/2021/09/001",
    journal = "JCAP",
    volume = "09",
    pages = "001",
    year = "2021"
}

@article{Quddus:2019ghy,
    author = "Quddus, Abdul and Panotopoulos, Grigorios and Kumar, Bharat and Ahmad, Shakeb and Patra, S. K.",
    title = "{GW170817 constraints on the properties of a neutron star in the presence of WIMP dark matter}",
    eprint = "1902.00929",
    archivePrefix = "arXiv",
    primaryClass = "nucl-th",
    doi = "10.1088/1361-6471/ab9d36",
    journal = "J. Phys. G",
    volume = "47",
    number = "9",
    pages = "095202",
    year = "2020"
}

@article{Miao:2022rqj,
    author = "Miao, Zhiqiang and Zhu, Yaofeng and Li, Ang and Huang, Feng",
    title = "{Dark Matter Admixed Neutron Star Properties in the Light of X-Ray Pulse Profile Observations}",
    eprint = "2204.05560",
    archivePrefix = "arXiv",
    primaryClass = "astro-ph.HE",
    doi = "10.3847/1538-4357/ac8544",
    journal = "Astrophys. J.",
    volume = "936",
    number = "1",
    pages = "69",
    year = "2022"
}

@article{Lopes:2023uxi,
    author = "Lopes, Luiz L. and Das, H. C.",
    title = "{Strange stars within bosonic and fermionic admixed dark matter}",
    eprint = "2301.00567",
    archivePrefix = "arXiv",
    primaryClass = "astro-ph.HE",
    doi = "10.1088/1475-7516/2023/05/034",
    journal = "JCAP",
    volume = "05",
    pages = "034",
    year = "2023"
}

@article{Thakur:2024ejl,
    author = "Thakur, Pratik and Kumar, Anil and Thapa, Vivek Baruah and Parmar, Vishal and Sinha, Monika",
    title = "{Exploring non-radial oscillation modes in dark matter admixed neutron stars}",
    eprint = "2406.07470",
    archivePrefix = "arXiv",
    primaryClass = "astro-ph.HE",
    doi = "10.1088/1475-7516/2024/12/042",
    journal = "JCAP",
    volume = "12",
    pages = "042",
    year = "2024"
}

@article{Lee:2021yyn,
    author = "Lee, Billy K. K. and Chu, Ming-chung and Lin, Lap-Ming",
    title = "{Could the GW190814 Secondary Component Be a Bosonic Dark Matter Admixed Compact Star?}",
    eprint = "2110.05538",
    archivePrefix = "arXiv",
    primaryClass = "astro-ph.HE",
    doi = "10.3847/1538-4357/ac2735",
    journal = "Astrophys. J.",
    volume = "922",
    number = "2",
    pages = "242",
    year = "2021"
}

@article{Gusakov:2005it,
    author = "Gusakov, M. E. and Kaminker, A. D. and Yakovlev, Dima G. and Gnedin, Oleg Y.",
    title = "{Cooling of Akmal-Pandharipande-Ravenhall neutron star models}",
    eprint = "astro-ph/0507560",
    archivePrefix = "arXiv",
    doi = "10.1111/j.1365-2966.2005.09459.x",
    journal = "Mon. Not. Roy. Astron. Soc.",
    volume = "363",
    pages = "555--562",
    year = "2005"
}

@article{Abac:2021txj,
    author = "Abac, Adrian G. and Bernido, Christopher C. and Esguerra, Jose Perico H.",
    title = "{Stability of neutron stars with dark matter core using three crustal types and the impact on mass{\textendash}radius relations}",
    eprint = "2104.04969",
    archivePrefix = "arXiv",
    primaryClass = "nucl-th",
    doi = "10.1016/j.dark.2023.101185",
    journal = "Phys. Dark Univ.",
    volume = "40",
    pages = "101185",
    year = "2023"
}

@article{Kain:2021hpk,
    author = "Kain, Ben",
    title = "{Dark matter admixed neutron stars}",
    eprint = "2102.08257",
    archivePrefix = "arXiv",
    primaryClass = "gr-qc",
    doi = "10.1103/PhysRevD.103.043009",
    journal = "Phys. Rev. D",
    volume = "103",
    number = "4",
    pages = "043009",
    year = "2021"
}

@article{DelPopolo:2020hel,
    author = "Del Popolo, Antonino and Le Delliou, Morgan and Deliyergiyev, Maksym",
    title = "{Neutron Stars and Dark Matter}",
    eprint = "2410.06078",
    archivePrefix = "arXiv",
    primaryClass = "astro-ph.CO",
    doi = "10.3390/universe6120222",
    journal = "Universe",
    volume = "6",
    number = "12",
    pages = "222",
    year = "2020"
}

@article{Kumar:2025ytm,
    author = "Kumar, Ankit and Sotani, Hajime",
    title = "{Impact of dark matter distribution on neutron star properties}",
    eprint = "2501.07052",
    archivePrefix = "arXiv",
    primaryClass = "astro-ph.HE",
    reportNumber = "RIKEN-iTHEMS-Report-25",
    doi = "10.1103/PhysRevD.111.043016",
    journal = "Phys. Rev. D",
    volume = "111",
    number = "4",
    pages = "043016",
    year = "2025"
}

@article{Das:2020ecp,
    author = "Das, Arpan and Malik, Tuhin and Nayak, Alekha C.",
    title = "{Dark matter admixed neutron star properties in light of gravitational wave observations: A two fluid approach}",
    eprint = "2011.01318",
    archivePrefix = "arXiv",
    primaryClass = "nucl-th",
    doi = "10.1103/PhysRevD.105.123034",
    journal = "Phys. Rev. D",
    volume = "105",
    number = "12",
    pages = "123034",
    year = "2022"
}

@article{Das:2018frc,
    author = "Das, Arpan and Malik, Tuhin and Nayak, Alekha C.",
    title = "{Confronting nuclear equation of state in the presence of dark matter using GW170817 observation in relativistic mean field theory approach}",
    eprint = "1807.10013",
    archivePrefix = "arXiv",
    primaryClass = "hep-ph",
    doi = "10.1103/PhysRevD.99.043016",
    journal = "Phys. Rev. D",
    volume = "99",
    number = "4",
    pages = "043016",
    year = "2019"
}

@article{Guha:2024pnn,
    author = "Guha, Atanu and Sen, Debashree",
    title = "{Constraining the mass of fermionic dark matter from its feeble interaction with hadronic matter via dark mediators in neutron stars}",
    eprint = "2401.14419",
    archivePrefix = "arXiv",
    primaryClass = "astro-ph.HE",
    doi = "10.1103/PhysRevD.109.043038",
    journal = "Phys. Rev. D",
    volume = "109",
    number = "4",
    pages = "043038",
    year = "2024"
}

@article{Chatterjee:2018prm,
    author = "Chatterjee, Debarati and Novak, Jerome and Oertel, Micaela",
    title = "{Magnetic field distribution in magnetars}",
    eprint = "1808.01778",
    archivePrefix = "arXiv",
    primaryClass = "nucl-th",
    doi = "10.1103/PhysRevC.99.055811",
    journal = "Phys. Rev. C",
    volume = "99",
    number = "5",
    pages = "055811",
    year = "2019"
}

@article{Tolman:1939jz,
    author = "Tolman, Richard C.",
    title = "{Static solutions of Einstein's field equations for spheres of fluid}",
    doi = "10.1103/PhysRev.55.364",
    journal = "Phys. Rev.",
    volume = "55",
    pages = "364--373",
    year = "1939"
}

@article{Huang:2020cab,
    author = "Huang, Kaixuan and Hu, Jinniu and Zhang, Ying and Shen, Hong",
    title = "{The possibility of the secondary object in GW190814 as a neutron star}",
    eprint = "2008.04491",
    archivePrefix = "arXiv",
    primaryClass = "nucl-th",
    doi = "10.3847/1538-4357/abbb37",
    journal = "Astrophys. J.",
    volume = "904",
    number = "1",
    pages = "39",
    year = "2020"
}

@article{Malik:2022zol,
    author = "Malik, Tuhin and Ferreira, M{\'a}rcio and Agrawal, B. K. and Provid{\^e}ncia, Constan{\c{c}}a",
    title = "{Relativistic Description of Dense Matter Equation of State and Compatibility with Neutron Star Observables: A Bayesian Approach}",
    eprint = "2201.12552",
    archivePrefix = "arXiv",
    primaryClass = "nucl-th",
    doi = "10.3847/1538-4357/ac5d3c",
    journal = "Astrophys. J.",
    volume = "930",
    number = "1",
    pages = "17",
    year = "2022"
}

@article{Char:2023fue,
    author = "Char, Prasanta and Mondal, Chiranjib and Gulminelli, Francesca and Oertel, Micaela",
    title = "{Generalized description of neutron star matter with a nucleonic relativistic density functional}",
    eprint = "2307.12364",
    archivePrefix = "arXiv",
    primaryClass = "nucl-th",
    doi = "10.1103/PhysRevD.108.103045",
    journal = "Phys. Rev. D",
    volume = "108",
    number = "10",
    pages = "103045",
    year = "2023"
}

@article{Riley:2019yda,
    author = "Riley, Thomas E. and others",
    title = "{A $NICER$ View of PSR J0030+0451: Millisecond Pulsar Parameter Estimation}",
    eprint = "1912.05702",
    archivePrefix = "arXiv",
    primaryClass = "astro-ph.HE",
    doi = "10.3847/2041-8213/ab481c",
    journal = "Astrophys. J. Lett.",
    volume = "887",
    number = "1",
    pages = "L21",
    year = "2019"
}

@article{Antoniadis:2013pzd,
    author = "Antoniadis, John and others",
    title = "{A Massive Pulsar in a Compact Relativistic Binary}",
    eprint = "1304.6875",
    archivePrefix = "arXiv",
    primaryClass = "astro-ph.HE",
    doi = "10.1126/science.1233232",
    journal = "Science",
    volume = "340",
    pages = "6131",
    year = "2013"
}

@article{Fonseca:2021wxt,
    author = "Fonseca, E. and others",
    title = "{Refined Mass and Geometric Measurements of the High-mass PSR J0740+6620}",
    eprint = "2104.00880",
    archivePrefix = "arXiv",
    primaryClass = "astro-ph.HE",
    doi = "10.3847/2041-8213/ac03b8",
    journal = "Astrophys. J. Lett.",
    volume = "915",
    number = "1",
    pages = "L12",
    year = "2021"
}

@article{Demorest:2010bx,
    author = "Demorest, Paul and Pennucci, Tim and Ransom, Scott and Roberts, Mallory and Hessels, Jason",
    title = "{Shapiro Delay Measurement of A Two Solar Mass Neutron Star}",
    eprint = "1010.5788",
    archivePrefix = "arXiv",
    primaryClass = "astro-ph.HE",
    doi = "10.1038/nature09466",
    journal = "Nature",
    volume = "467",
    pages = "1081--1083",
    year = "2010"
}

@article{Romani:2022jhd,
    author = "Romani, Roger W. and Kandel, D. and Filippenko, Alexei V. and Brink, Thomas G. and Zheng, WeiKang",
    title = "{PSR J0952{\ensuremath{-}}0607: The Fastest and Heaviest Known Galactic Neutron Star}",
    eprint = "2207.05124",
    archivePrefix = "arXiv",
    primaryClass = "astro-ph.HE",
    doi = "10.3847/2041-8213/ac8007",
    journal = "Astrophys. J. Lett.",
    volume = "934",
    number = "2",
    pages = "L17",
    year = "2022"
}

@article{LIGOScientific:2018cki,
    author = "Abbott, B. P. and others",
    collaboration = "LIGO Scientific, Virgo",
    title = "{GW170817: Measurements of neutron star radii and equation of state}",
    eprint = "1805.11581",
    archivePrefix = "arXiv",
    primaryClass = "gr-qc",
    reportNumber = "LIGO-P1800115",
    doi = "10.1103/PhysRevLett.121.161101",
    journal = "Phys. Rev. Lett.",
    volume = "121",
    number = "16",
    pages = "161101",
    year = "2018"
}

@article{Bell:2020jou,
    author = "Bell, Nicole F. and Busoni, Giorgio and Robles, Sandra and Virgato, Michael",
    title = "{Improved Treatment of Dark Matter Capture in Neutron Stars}",
    eprint = "2004.14888",
    archivePrefix = "arXiv",
    primaryClass = "hep-ph",
    doi = "10.1088/1475-7516/2020/09/028",
    journal = "JCAP",
    volume = "09",
    pages = "028",
    year = "2020"
}

@article{Das:2021dru,
    author = "Das, H. C. and Kumar, Ankit and Biswal, S. K. and Patra, S. K.",
    title = "{Impacts of dark matter on the f-mode oscillation of hyperon star}",
    eprint = "2109.01851",
    archivePrefix = "arXiv",
    primaryClass = "nucl-th",
    doi = "10.1103/PhysRevD.104.123006",
    journal = "Phys. Rev. D",
    volume = "104",
    number = "12",
    pages = "123006",
    year = "2021"
}

@article{Kumar:2024zzl,
    author = "Kumar, Ankit and Sotani, Hajime",
    title = "{Constraints on the parameter space in dark matter admixed neutron stars}",
    eprint = "2408.15312",
    archivePrefix = "arXiv",
    primaryClass = "astro-ph.HE",
    reportNumber = "RIKEN-iTHEMS-Report-24",
    doi = "10.1103/PhysRevD.110.063001",
    journal = "Phys. Rev. D",
    volume = "110",
    number = "6",
    pages = "063001",
    year = "2024"
}

@article{Passarella:2025zqb,
    author = "Passarella, Luca and Margueron, Jerome and Pagliara, Giuseppe",
    title = "{Relativistic mean-field predictions for the dense-matter equation~of state and application to neutron stars}",
    eprint = "2503.23028",
    archivePrefix = "arXiv",
    primaryClass = "nucl-th",
    doi = "10.1103/7qs4-wb95",
    journal = "Phys. Rev. C",
    volume = "112",
    number = "3",
    pages = "035805",
    year = "2025"
}

@article{Jyothilakshmi:2024xtl,
    author = "Jyothilakshmi, O. P. and Naik, Lakshmi J. and Sen, Debashree and Guha, Atanu and Sreekanth, V.",
    title = "{Effects of dark boson mediated feeble interaction between dark matter (DM) and quark matter on f-mode oscillation of DM admixed quark stars}",
    eprint = "2410.20923",
    archivePrefix = "arXiv",
    primaryClass = "hep-ph",
    doi = "10.1140/epjc/s10052-025-14109-w",
    journal = "Eur. Phys. J. C",
    volume = "85",
    number = "4",
    pages = "461",
    year = "2025"
}

@article{Kumar:2022amh,
    author = "Kumar, Ankit and Das, H. C. and Patra, S. K.",
    title = "{Thermal relaxation of dark matter admixed neutron star}",
    eprint = "2203.02132",
    archivePrefix = "arXiv",
    primaryClass = "astro-ph.HE",
    doi = "10.1093/mnras/stac1013",
    journal = "Mon. Not. Roy. Astron. Soc.",
    volume = "513",
    number = "2",
    pages = "1820--1833",
    year = "2022"
}

@article{Kovetz:2017rvv,
    author = "Kovetz, Ely D.",
    title = "{Probing Primordial-Black-Hole Dark Matter with Gravitational Waves}",
    eprint = "1705.09182",
    archivePrefix = "arXiv",
    primaryClass = "astro-ph.CO",
    doi = "10.1103/PhysRevLett.119.131301",
    journal = "Phys. Rev. Lett.",
    volume = "119",
    number = "13",
    pages = "131301",
    year = "2017"
}

@article{Loc:2024qbz,
    author = "Loc, Ngo Phuc Duc",
    title = "{Gravitational waves from burdened primordial black holes dark matter}",
    eprint = "2410.17544",
    archivePrefix = "arXiv",
    primaryClass = "gr-qc",
    doi = "10.1103/PhysRevD.111.023509",
    journal = "Phys. Rev. D",
    volume = "111",
    number = "2",
    pages = "023509",
    year = "2025"
}

@article{Dasgupta:2019cae,
    author = "Dasgupta, Basudeb and Laha, Ranjan and Ray, Anupam",
    title = "{Neutrino and positron constraints on spinning primordial black hole dark matter}",
    eprint = "1912.01014",
    archivePrefix = "arXiv",
    primaryClass = "hep-ph",
    reportNumber = "CERN-TH-2019-212, TIFR/TH/19-40",
    doi = "10.1103/PhysRevLett.125.101101",
    journal = "Phys. Rev. Lett.",
    volume = "125",
    number = "10",
    pages = "101101",
    year = "2020"
}

@article{Carr:2016drx,
    author = "Carr, Bernard and Kuhnel, Florian and Sandstad, Marit",
    title = "{Primordial Black Holes as Dark Matter}",
    eprint = "1607.06077",
    archivePrefix = "arXiv",
    primaryClass = "astro-ph.CO",
    reportNumber = "NORDITA-2016-83",
    doi = "10.1103/PhysRevD.94.083504",
    journal = "Phys. Rev. D",
    volume = "94",
    number = "8",
    pages = "083504",
    year = "2016"
}

@article{Iguaz:2021irx,
    author = "Iguaz, J. and Iguaz, J. and Serpico, P. D. and Serpico, P. D. and Siegert, T. and Siegert, T.",
    title = "{Isotropic X-ray bound on Primordial Black Hole Dark Matter}",
    eprint = "2104.03145",
    archivePrefix = "arXiv",
    primaryClass = "astro-ph.CO",
    reportNumber = "LAPTH-014/21",
    doi = "10.1103/PhysRevD.103.103025",
    journal = "Phys. Rev. D",
    volume = "103",
    number = "10",
    pages = "103025",
    year = "2021",
    note = "[Erratum: Phys.Rev.D 107, 069902 (2023)]"
}

@article{Xie:2024eug,
    author = "Xie, Zhen and Liu, Bing and Liu, Jiahao and Cai, Yi-Fu and Yang, Ruizhi",
    title = "{Limits on the primordial black holes dark matter with future MeV detectors}",
    eprint = "2401.06440",
    archivePrefix = "arXiv",
    primaryClass = "astro-ph.HE",
    doi = "10.1103/PhysRevD.109.043020",
    journal = "Phys. Rev. D",
    volume = "109",
    number = "4",
    pages = "043020",
    year = "2024"
}

@article{Jedamzik:2020omx,
    author = "Jedamzik, Karsten",
    title = "{Consistency of Primordial Black Hole Dark Matter with LIGO/Virgo Merger Rates}",
    eprint = "2007.03565",
    archivePrefix = "arXiv",
    primaryClass = "astro-ph.CO",
    doi = "10.1103/PhysRevLett.126.051302",
    journal = "Phys. Rev. Lett.",
    volume = "126",
    number = "5",
    pages = "051302",
    year = "2021"
}

@article{Kouvaris:2007ay,
    author = "Kouvaris, Chris",
    title = "{WIMP Annihilation and Cooling of Neutron Stars}",
    eprint = "0708.2362",
    archivePrefix = "arXiv",
    primaryClass = "astro-ph",
    doi = "10.1103/PhysRevD.77.023006",
    journal = "Phys. Rev. D",
    volume = "77",
    pages = "023006",
    year = "2008"
}

@article{Lecce:2025dbz,
    author = "Lecce, Francesca and Lella, Alessandro and Lucente, Giuseppe and Vijayan, Vimal and Bauswein, Andreas and Giannotti, Maurizio and Mirizzi, Alessandro",
    title = "{Probing axionlike particles with multimessenger observations of neutron star mergers}",
    eprint = "2504.02032",
    archivePrefix = "arXiv",
    primaryClass = "hep-ph",
    reportNumber = "BARI-TH/773-25",
    doi = "10.1103/krf3-lm4s",
    journal = "Phys. Rev. D",
    volume = "112",
    number = "2",
    pages = "023001",
    year = "2025"
}

@article{Beznogov:2018fda,
    author = "Beznogov, Mikhail V. and Rrapaj, Ermal and Page, Dany and Reddy, Sanjay",
    title = "{Constraints on Axion-like Particles and Nucleon Pairing in Dense Matter from the Hot Neutron Star in HESS J1731-347}",
    eprint = "1806.07991",
    archivePrefix = "arXiv",
    primaryClass = "astro-ph.HE",
    reportNumber = "INT-PUB-18-029",
    doi = "10.1103/PhysRevC.98.035802",
    journal = "Phys. Rev. C",
    volume = "98",
    number = "3",
    pages = "035802",
    year = "2018"
}

@article{Fiorillo:2022piv,
    author = "Fiorillo, Damiano F. G. and Iocco, Fabio",
    title = "{Axions from neutron star mergers}",
    doi = "10.1103/PhysRevD.105.123007",
    journal = "Phys. Rev. D",
    volume = "105",
    number = "12",
    pages = "123007",
    year = "2022"
}

@article{Dev:2023hax,
    author = "Dev, P. S. Bhupal and Fortin, Jean-Fran{\c{c}}ois and Harris, Steven P. and Sinha, Kuver and Zhang, Yongchao",
    title = "{First Constraints on the Photon Coupling of Axionlike Particles from Multimessenger Studies of the Neutron Star Merger GW170817}",
    eprint = "2305.01002",
    archivePrefix = "arXiv",
    primaryClass = "hep-ph",
    reportNumber = "INT-PUB-23-014",
    doi = "10.1103/PhysRevLett.132.101003",
    journal = "Phys. Rev. Lett.",
    volume = "132",
    number = "10",
    pages = "101003",
    year = "2024"
}

@article{Bhusal:2020bvx,
    author = "Bhusal, Aagaman and Houston, Nick and Li, Tianjun",
    title = "{Searching for Solar Axions Using Data from the Sudbury Neutrino Observatory}",
    eprint = "2004.02733",
    archivePrefix = "arXiv",
    primaryClass = "hep-ph",
    doi = "10.1103/PhysRevLett.126.091601",
    journal = "Phys. Rev. Lett.",
    volume = "126",
    number = "9",
    pages = "091601",
    year = "2021"
}

@article{Sedrakian:2015krq,
    author = "Sedrakian, Armen",
    title = "{Axion cooling of neutron stars}",
    eprint = "1512.07828",
    archivePrefix = "arXiv",
    primaryClass = "astro-ph.HE",
    doi = "10.1103/PhysRevD.93.065044",
    journal = "Phys. Rev. D",
    volume = "93",
    number = "6",
    pages = "065044",
    year = "2016"
}

@article{Zhang:2023vva,
    author = "Zhang, Hong-Yi and Hagimoto, Ray and Long, Andrew J.",
    title = "{Neutron star cooling with lepton-flavor-violating axions}",
    eprint = "2309.03889",
    archivePrefix = "arXiv",
    primaryClass = "hep-ph",
    doi = "10.1103/PhysRevD.109.103005",
    journal = "Phys. Rev. D",
    volume = "109",
    number = "10",
    pages = "103005",
    year = "2024"
}

@article{Rutherford:2022xeb,
    author = "Rutherford, Nathan and Raaijmakers, Geert and Prescod-Weinstein, Chanda and Watts, Anna",
    title = "{Constraining bosonic asymmetric dark matter with neutron star mass-radius measurements}",
    eprint = "2208.03282",
    archivePrefix = "arXiv",
    primaryClass = "astro-ph.HE",
    doi = "10.1103/PhysRevD.107.103051",
    journal = "Phys. Rev. D",
    volume = "107",
    number = "10",
    pages = "103051",
    year = "2023"
}

@article{Sun:2023cqr,
    author = "Sun, Hongyi and Wen, Dehua",
    title = "{New criterion for the existence of dark matter in neutron stars}",
    eprint = "2312.17288",
    archivePrefix = "arXiv",
    primaryClass = "astro-ph.HE",
    doi = "10.1103/PhysRevD.109.123037",
    journal = "Phys. Rev. D",
    volume = "109",
    number = "12",
    pages = "123037",
    year = "2024"
}

@article{Pitz:2024xvh,
    author = {Pitz, Sarah Louisa and Schaffner-Bielich, J{\"u}rgen},
    title = "{Generating ultracompact neutron stars with bosonic dark matter}",
    eprint = "2408.13157",
    archivePrefix = "arXiv",
    primaryClass = "astro-ph.HE",
    doi = "10.1103/PhysRevD.111.043050",
    journal = "Phys. Rev. D",
    volume = "111",
    number = "4",
    pages = "043050",
    year = "2025"
}

@article{Das:2021yny,
    author = "Das, H. C. and Kumar, Ankit and Patra, S. K.",
    title = "{Dark matter admixed neutron star as a possible compact component in the GW190814 merger event}",
    eprint = "2109.01853",
    archivePrefix = "arXiv",
    primaryClass = "astro-ph.HE",
    doi = "10.1103/PhysRevD.104.063028",
    journal = "Phys. Rev. D",
    volume = "104",
    number = "6",
    pages = "063028",
    year = "2021"
}

@article{Liu:2024rix,
    author = "Liu, Hong-Ming and Wei, Jin-Biao and Li, Zeng-Hua and Burgio, G. F. and Das, H. C. and Schulze, H. -J.",
    title = "{Dark matter effects on the properties of neutron stars: Compactness and tidal deformability}",
    eprint = "2403.17024",
    archivePrefix = "arXiv",
    primaryClass = "nucl-th",
    doi = "10.1103/PhysRevD.110.023024",
    journal = "Phys. Rev. D",
    volume = "110",
    number = "2",
    pages = "023024",
    year = "2024"
}

@article{Routaray:2022utr,
    author = "Routaray, Pinku and Das, H. C. and Sen, Souhardya and Kumar, Bharat and Panotopoulos, Grigoris and Zhao, Tianqi",
    title = "{Radial oscillations of dark matter admixed neutron stars}",
    eprint = "2211.12808",
    archivePrefix = "arXiv",
    primaryClass = "nucl-th",
    doi = "10.1103/PhysRevD.107.103039",
    journal = "Phys. Rev. D",
    volume = "107",
    number = "10",
    pages = "103039",
    year = "2023"
}

@article{Kumar:2025cro,
    author = "Kumar, Ankit and Sotani, Hajime",
    title = "{Asteroseismology and universal relations in neutron stars with gravitationally bound dark matter}",
    eprint = "2506.00311",
    archivePrefix = "arXiv",
    primaryClass = "astro-ph.HE",
    reportNumber = "RIKEN-iTHEMS-Report-25",
    doi = "10.1103/l9p9-nqbx",
    journal = "Phys. Rev. D",
    volume = "111",
    number = "12",
    pages = "123028",
    year = "2025"
}

@article{Sen:2024yim,
    author = "Sen, Debashree and Guha, Atanu",
    title = "{Impact of a dark boson mediated feeble interaction between dark matter and hadronic matter on the f-mode oscillation of neutron stars}",
    eprint = "2409.18890",
    archivePrefix = "arXiv",
    primaryClass = "hep-ph",
    doi = "10.1103/PhysRevD.110.103013",
    journal = "Phys. Rev. D",
    volume = "110",
    number = "10",
    pages = "103013",
    year = "2024"
}

@article{Acevedo:2024ttq,
    author = "Acevedo, Javier F. and Bramante, Joseph and Liu, Qinrui and Tyagi, Narayani",
    title = "{Neutrino and gamma-ray signatures of inelastic dark matter annihilating outside neutron stars}",
    eprint = "2404.10039",
    archivePrefix = "arXiv",
    primaryClass = "hep-ph",
    reportNumber = "SLAC-PUB-17767",
    doi = "10.1088/1475-7516/2025/03/028",
    journal = "JCAP",
    volume = "03",
    pages = "028",
    year = "2025"
}

@article{Bell:2023ysh,
    author = "Bell, Nicole F. and Busoni, Giorgio and Robles, Sandra and Virgato, Michael",
    title = "{Thermalization and annihilation of dark matter in neutron stars}",
    eprint = "2312.11892",
    archivePrefix = "arXiv",
    primaryClass = "hep-ph",
    reportNumber = "KCL-PH-TH/2023-71, FERMILAB-PUB-23-810-T",
    doi = "10.1088/1475-7516/2024/04/006",
    journal = "JCAP",
    volume = "04",
    pages = "006",
    year = "2024"
}

@article{Pal:2024afl,
    author = "Pal, Suman and Chaudhuri, Gargi",
    title = "{Effect of dark matter interaction on hybrid star in the light of the recent astrophysical observations}",
    eprint = "2405.04856",
    archivePrefix = "arXiv",
    primaryClass = "nucl-th",
    doi = "10.1088/1475-7516/2024/10/064",
    journal = "JCAP",
    volume = "10",
    pages = "064",
    year = "2024"
}

@article{Shahrbaf:2025hsw,
    author = "Shahrbaf, Mahboubeh and Thakur, Prashant and Rafiei Karkevandi, Davood",
    title = "{Probing strange dark matter through f-mode oscillations of neutron stars with hyperons and quark matter}",
    eprint = "2510.08115",
    archivePrefix = "arXiv",
    primaryClass = "nucl-th",
    doi = "10.1088/1475-7516/2026/03/017",
    journal = "JCAP",
    volume = "03",
    pages = "017",
    year = "2026"
}

@article{Cermeno:2017ejm,
    author = "Cerme{\~n}o, M. and P{\'e}rez-Garc{\'\i}a, M. A. and Lineros, R. A.",
    title = "{Enhanced neutrino emissivities in pseudoscalar-mediated Dark Matter annihilation in Neutron Stars}",
    eprint = "1705.03012",
    archivePrefix = "arXiv",
    primaryClass = "hep-ph",
    doi = "10.3847/1538-4357/aad1ec",
    journal = "Astrophys. J.",
    volume = "863",
    pages = "157",
    year = "2018"
}

@article{Cline:2013gha,
    author = "Cline, James M. and Kainulainen, Kimmo and Scott, Pat and Weniger, Christoph",
    title = "{Update on scalar singlet dark matter}",
    eprint = "1306.4710",
    archivePrefix = "arXiv",
    primaryClass = "hep-ph",
    doi = "10.1103/PhysRevD.88.055025",
    journal = "Phys. Rev. D",
    volume = "88",
    pages = "055025",
    year = "2013",
    note = "[Erratum: Phys.Rev.D 92, 039906 (2015)]"
}

@article{Parmar:2023zlg,
    author = "Parmar, Vishal and Das, H. C. and Sharma, M. K. and Patra, S. K.",
    title = "{Influence of dark matter on magnetized neutron stars}",
    eprint = "2306.17510",
    archivePrefix = "arXiv",
    primaryClass = "astro-ph.HE",
    doi = "10.1103/PhysRevD.108.083003",
    journal = "Phys. Rev. D",
    volume = "108",
    number = "8",
    pages = "083003",
    year = "2023"
}

@article{Routaray:2024fcq,
    author = "Routaray, Pinku and Parmar, Vishal and Das, H. C. and Kumar, Bharat and Burgio, G. F. and Schulze, H. -J.",
    title = "{Effects of asymmetric dark matter on a magnetized neutron star: A two-fluid approach}",
    eprint = "2412.21097",
    archivePrefix = "arXiv",
    primaryClass = "nucl-th",
    doi = "10.1103/PhysRevD.111.103045",
    journal = "Phys. Rev. D",
    volume = "111",
    number = "10",
    pages = "103045",
    year = "2025"
}

@article{Kaspi:2017fwg,
    author = "Kaspi, Victoria M. and Beloborodov, Andrei",
    title = "{Magnetars}",
    eprint = "1703.00068",
    archivePrefix = "arXiv",
    primaryClass = "astro-ph.HE",
    doi = "10.1146/annurev-astro-081915-023329",
    journal = "Ann. Rev. Astron. Astrophys.",
    volume = "55",
    pages = "261--301",
    year = "2017"
}

@article{Avancini:2017gck,
    author = "Avancini, Sidney S. and Dexheimer, Veronica and Farias, Ricardo L. S. and Tim{\'o}teo, Varese S.",
    title = "{Anisotropy in the equation of state of strongly magnetized quark matter within the Nambu{\textendash}Jona-Lasinio model}",
    eprint = "1709.02774",
    archivePrefix = "arXiv",
    primaryClass = "hep-ph",
    doi = "10.1103/PhysRevC.97.035207",
    journal = "Phys. Rev. C",
    volume = "97",
    number = "3",
    pages = "035207",
    year = "2018"
}

@article{Tolos:2016hhl,
    author = "Tolos, Laura and Centelles, Mario and Ramos, Angels",
    title = "{Equation of State for Nucleonic and Hyperonic Neutron Stars with Mass and Radius Constraints}",
    eprint = "1610.00919",
    archivePrefix = "arXiv",
    primaryClass = "astro-ph.HE",
    doi = "10.3847/1538-4357/834/1/3",
    journal = "Astrophys. J.",
    volume = "834",
    number = "1",
    pages = "3",
    year = "2017"
}

@article{Franzon:2016iai,
    author = "Franzon, B. and Dexheimer, V. and Schramm, S.",
    title = "{Internal composition of proto-neutron stars under strong magnetic fields}",
    eprint = "1606.04843",
    archivePrefix = "arXiv",
    primaryClass = "astro-ph.HE",
    doi = "10.1103/PhysRevD.94.044018",
    journal = "Phys. Rev. D",
    volume = "94",
    number = "4",
    pages = "044018",
    year = "2016"
}

@article{Franzon:2015sya,
    author = "Franzon, B. and Dexheimer, V. and Schramm, S.",
    title = "{A self-consistent study of magnetic field effects on hybrid stars}",
    eprint = "1508.04431",
    archivePrefix = "arXiv",
    primaryClass = "astro-ph.HE",
    doi = "10.1093/mnras/stv2606",
    journal = "Mon. Not. Roy. Astron. Soc.",
    volume = "456",
    number = "3",
    pages = "2937--2945",
    year = "2016"
}

@article{Chatterjee:2014qsa,
    author = "Chatterjee, Debarati and Elghozi, Thomas and Novak, Jerome and Oertel, Micaela",
    title = "{Consistent neutron star models with magnetic field dependent equations of state}",
    eprint = "1410.6332",
    archivePrefix = "arXiv",
    primaryClass = "astro-ph.HE",
    doi = "10.1093/mnras/stu2706",
    journal = "Mon. Not. Roy. Astron. Soc.",
    volume = "447",
    pages = "3785",
    year = "2015"
}

@article{Bowers:1974tgi,
    author = "Bowers, Richard L. and Liang, E. P. T.",
    title = "{Anisotropic Spheres in General Relativity}",
    doi = "10.1086/152760",
    journal = "Astrophys. J.",
    volume = "188",
    pages = "657--665",
    year = "1974"
}

@article{Bonazzola:1993zz,
    author = "Bonazzola, S. and Gourgoulhon, E. and Salgado, M. and Marck, J. A.",
    title = "{Axisymmetric rotating relativistic bodies: A new numerical approach for 'exact' solutions}",
    journal = "Astron. Astrophys.",
    volume = "278",
    pages = "421--443",
    year = "1993"
}

@article{Broderick:2000pe,
    author = "Broderick, A. and Prakash, M. and Lattimer, J. M.",
    title = "{The Equation of state of neutron star matter in strong magnetic fields}",
    eprint = "astro-ph/0001537",
    archivePrefix = "arXiv",
    doi = "10.1086/309010",
    journal = "Astrophys. J.",
    volume = "537",
    pages = "351",
    year = "2000"
}

@article{Strickland:2012vu,
    author = "Strickland, M. and Dexheimer, V. and Menezes, D. P.",
    title = "{Bulk Properties of a Fermi Gas in a Magnetic Field}",
    eprint = "1209.3276",
    archivePrefix = "arXiv",
    primaryClass = "nucl-th",
    doi = "10.1103/PhysRevD.86.125032",
    journal = "Phys. Rev. D",
    volume = "86",
    pages = "125032",
    year = "2012"
}

@article{Boguta:1977xi,
    author = "Boguta, J. and Bodmer, A. R.",
    title = "{Relativistic Calculation of Nuclear Matter and the Nuclear Surface}",
    doi = "10.1016/0375-9474(77)90626-1",
    journal = "Nucl. Phys. A",
    volume = "292",
    pages = "413--428",
    year = "1977"
}

@article{Mueller:1996pm,
    author = "Mueller, Horst and Serot, Brian D.",
    title = "{Relativistic mean field theory and the high density nuclear equation of state}",
    eprint = "nucl-th/9603037",
    archivePrefix = "arXiv",
    reportNumber = "IU-NTC-96-03",
    doi = "10.1016/0375-9474(96)00187-X",
    journal = "Nucl. Phys. A",
    volume = "606",
    pages = "508--537",
    year = "1996"
}

@article{Steiner:2004fi,
    author = "Steiner, Andrew W. and Prakash, Madappa and Lattimer, James M. and Ellis, Paul J.",
    title = "{Isospin asymmetry in nuclei and neutron stars}",
    eprint = "nucl-th/0410066",
    archivePrefix = "arXiv",
    reportNumber = "LA-UR-04-6745",
    doi = "10.1016/j.physrep.2005.02.004",
    journal = "Phys. Rept.",
    volume = "411",
    pages = "325--375",
    year = "2005"
}

@article{Todd-Rutel:2005yzo,
    author = "Todd-Rutel, B. G. and Piekarewicz, J.",
    title = "{Neutron-Rich Nuclei and Neutron Stars: A New Accurately Calibrated Interaction for the Study of Neutron-Rich Matter}",
    eprint = "nucl-th/0504034",
    archivePrefix = "arXiv",
    doi = "10.1103/PhysRevLett.95.122501",
    journal = "Phys. Rev. Lett.",
    volume = "95",
    pages = "122501",
    year = "2005"
}

@article{Typel:1999yq,
    author = "Typel, S. and Wolter, H. H.",
    title = "{Relativistic mean field calculations with density dependent meson nucleon coupling}",
    doi = "10.1016/S0375-9474(99)00310-3",
    journal = "Nucl. Phys. A",
    volume = "656",
    pages = "331--364",
    year = "1999"
}

@article{Typel:2009sy,
    author = "Typel, S. and Ropke, G. and Klahn, T. and Blaschke, D. and Wolter, H. H.",
    title = "{Composition and thermodynamics of nuclear matter with light clusters}",
    eprint = "0908.2344",
    archivePrefix = "arXiv",
    primaryClass = "nucl-th",
    doi = "10.1103/PhysRevC.81.015803",
    journal = "Phys. Rev. C",
    volume = "81",
    pages = "015803",
    year = "2010"
}

@article{Lalazissis:2005de,
    author = "Lalazissis, G. A. and Niksic, T. and Vretenar, D. and Ring, P.",
    title = "{New relativistic mean-field interaction with density-dependent meson-nucleon couplings}",
    doi = "10.1103/PhysRevC.71.024312",
    journal = "Phys. Rev. C",
    volume = "71",
    pages = "024312",
    year = "2005"
}

@article{Patra:2020wjy,
    author = "Patra, N. K. and Malik, Tuhin and Sen, Debashree and Jha, T. K. and Mishra, Hiranmaya",
    title = "{An Equation of State for Magnetized Neutron Star Matter and Tidal Deformation in Neutron Star Mergers}",
    doi = "10.3847/1538-4357/aba8fc",
    journal = "Astrophys. J.",
    volume = "900",
    number = "1",
    pages = "49",
    year = "2020"
}

@article{Hinderer:2007mb,
    author = "Hinderer, Tanja",
    title = "{Tidal Love numbers of neutron stars}",
    eprint = "0711.2420",
    archivePrefix = "arXiv",
    primaryClass = "astro-ph",
    doi = "10.1086/533487",
    journal = "Astrophys. J.",
    volume = "677",
    pages = "1216--1220",
    year = "2008",
    note = "[Erratum: Astrophys.J. 697, 964 (2009)]"
}

@article{Panotopoulos:2017idn,
    author = "Panotopoulos, Grigorios and Lopes, Il{\'\i}dio",
    title = "{Dark matter effect on realistic equation of state in neutron stars}",
    eprint = "1709.06312",
    archivePrefix = "arXiv",
    primaryClass = "hep-ph",
    doi = "10.1103/PhysRevD.96.083004",
    journal = "Phys. Rev. D",
    volume = "96",
    number = "8",
    pages = "083004",
    year = "2017"
}

@article{deBlok:2001hbg,
    author = "de Blok, W. J. G. and McGaugh, Stacy S. and Bosma, Albert and Rubin, Vera C.",
    title = "{Mass density profiles of LSB galaxies}",
    eprint = "astro-ph/0103102",
    archivePrefix = "arXiv",
    doi = "10.1086/320262",
    journal = "Astrophys. J. Lett.",
    volume = "552",
    pages = "L23--L26",
    year = "2001"
}

@article{Bertone:2004pz,
    author = "Bertone, Gianfranco and Hooper, Dan and Silk, Joseph",
    title = "{Particle dark matter: Evidence, candidates and constraints}",
    eprint = "hep-ph/0404175",
    archivePrefix = "arXiv",
    reportNumber = "FERMILAB-PUB-04-047-A",
    doi = "10.1016/j.physrep.2004.08.031",
    journal = "Phys. Rept.",
    volume = "405",
    pages = "279--390",
    year = "2005"
}

@book{Bauer:2017qwy,
    author = "Bauer, Martin and Plehn, Tilman",
    title = "{Yet Another Introduction to Dark Matter}: {The Particle Physics Approach}",
    eprint = "1705.01987",
    archivePrefix = "arXiv",
    primaryClass = "hep-ph",
    doi = "10.1007/978-3-030-16234-4",
    publisher = "Springer",
    series = "Lecture Notes in Physics",
    volume = "959",
    year = "2019"
}

@article{Issifu:2025jac,
    author = "Issifu, Adamu and Thakur, Prashant and Rafiei Karkevandi, Davood and da Silva, Franciele M. and Menezes, D{\'e}bora P. and Lim, Y. and Frederico, Tobias",
    title = "{Dark matter heating in evolving protoneutron stars: A two-fluid approach}",
    eprint = "2511.07567",
    archivePrefix = "arXiv",
    primaryClass = "astro-ph.HE",
    doi = "10.1103/bhm3-jzjq",
    journal = "Phys. Rev. D",
    volume = "113",
    number = "10",
    pages = "103042",
    year = "2026"
}

@article{Yadav:2022yqa,
    author = "Yadav, Shubham and Mishra, M. and Sarkar, Tapomoy Guha and Singh, Captain R.",
    title = "{Thermal evolution and axion emission properties of strongly magnetized neutron stars}",
    eprint = "2212.11652",
    archivePrefix = "arXiv",
    primaryClass = "astro-ph.HE",
    doi = "10.1140/epjc/s10052-024-12583-2",
    journal = "Eur. Phys. J. C",
    volume = "84",
    number = "3",
    pages = "225",
    year = "2024"
}

@article{Huang:2023grj,
    author = "Huang, Chun and Raaijmakers, Geert and Watts, Anna L. and Tolos, Laura and Provid{\^e}ncia, Constan{\c{c}}a",
    title = "{Constraining a relativistic mean field model using neutron star mass{\textendash}radius measurements I: nucleonic models}",
    eprint = "2303.17518",
    archivePrefix = "arXiv",
    primaryClass = "astro-ph.HE",
    doi = "10.1093/mnras/stae844",
    journal = "Mon. Not. Roy. Astron. Soc.",
    volume = "529",
    number = "4",
    pages = "4650--4665",
    year = "2024"
}

@article{Huang:2024ewv,
    author = "Huang, Chun and Zheng, Xiao-Ping",
    title = "{Bayesian insights into post-glitch dynamics: model comparison and parameter constraint from decades long observation data of the Crab pulsar}",
    eprint = "2409.18432",
    archivePrefix = "arXiv",
    primaryClass = "astro-ph.HE",
    doi = "10.1093/mnras/staf1415",
    journal = "Mon. Not. Roy. Astron. Soc.",
    volume = "542",
    number = "4",
    pages = "3198--3205",
    year = "2025"
}

@article{Huang:2024wig,
    author = "Huang, Chun",
    title = "{Equation of State Independent Determination on the Radius of a 1.4 M$_{\odot}$ Neutron Star Using Mass{\textendash}Radius Measurements}",
    eprint = "2412.10242",
    archivePrefix = "arXiv",
    primaryClass = "astro-ph.HE",
    doi = "10.3847/2041-8213/ad9f3c",
    journal = "Astrophys. J. Lett.",
    volume = "978",
    number = "1",
    pages = "L14",
    year = "2025"
}

@article{Oppenheimer:1939ne,
    author = "Oppenheimer, J. R. and Volkoff, G. M.",
    title = "{On massive neutron cores}",
    doi = "10.1103/PhysRev.55.374",
    journal = "Phys. Rev.",
    volume = "55",
    pages = "374--381",
    year = "1939"
}

@article{Baym:1971pw,
    author = "Baym, Gordon and Pethick, Christopher and Sutherland, Peter",
    title = "{The Ground state of matter at high densities: Equation of state and stellar models}",
    doi = "10.1086/151216",
    journal = "Astrophys. J.",
    volume = "170",
    pages = "299--317",
    year = "1971"
}

@article{Hinderer:2009ca,
    author = "Hinderer, Tanja and Lackey, Benjamin D. and Lang, Ryan N. and Read, Jocelyn S.",
    title = "{Tidal deformability of neutron stars with realistic equations of state and their gravitational wave signatures in binary inspiral}",
    eprint = "0911.3535",
    archivePrefix = "arXiv",
    primaryClass = "astro-ph.HE",
    doi = "10.1103/PhysRevD.81.123016",
    journal = "Phys. Rev. D",
    volume = "81",
    pages = "123016",
    year = "2010"
}
\end{document}